\documentclass[review,onefignum,onetabnum]{article}

\usepackage{xspace}
\usepackage{amsfonts}
\usepackage{amsmath}
\usepackage{amssymb}
\usepackage{amsbsy}
\usepackage{bm}
\usepackage{cite}
\usepackage{color}
\usepackage{dsfont}
\usepackage{epsfig}
\usepackage{eucal}
\usepackage{graphicx}
\usepackage{mathrsfs}
\usepackage{multirow} 
\usepackage{tikz}

\usepackage{enumitem}
\setlist[enumerate]{leftmargin=0.35in}
\setlist[itemize]{leftmargin=0.35in}

\newcommand{\bi}{\begin{itemize}}
\newcommand{\ei}{\end{itemize}}

\def\vlam{{\bm \lambda}}
\def\vzeta{{\bm \zeta}}
\def\vtheta{{\bm \theta}}
\def\vxi{{\bm \xi}}

\def\vc{{\bm c}}
\def\vt{{\bm t}}
\def\vx{{\bm x}}
\def\vy{\bm y}
\def\vz{{\bm z}}

\usepackage{caption}
\newcommand{\bt}[1]{\textcolor{black}{#1}}

\RequirePackage{titlesec}
\usepackage{authblk}

\begin{document}

\centerline{\large \bf Polynomial Chaos Surrogate Construction for Random Fields with Parametric Uncertainty}
\medskip
\centerline{J.N.~Mueller, K.~Sargsyan, C.J.~Daniels, and H.N.~Najm}
\centerline{Sandia National Laboratories, Livermore, CA}
\centerline{\today}
\medskip

\begin{abstract}
\noindent Engineering and applied science rely on computational experiments to rigorously study physical systems. The mathematical models used to probe these systems are highly complex, and sampling-intensive studies often require prohibitively many simulations for acceptable accuracy. Surrogate models provide a means of circumventing the high computational expense of sampling such complex models. In particular, polynomial chaos expansions (PCEs) have been successfully used for uncertainty quantification studies of deterministic models where the dominant source of uncertainty is parametric. We discuss an extension to conventional PCE surrogate modeling to enable surrogate construction for stochastic computational models that have intrinsic noise in addition to parametric uncertainty. We develop a PCE surrogate on a joint space of intrinsic and parametric uncertainty, enabled by Rosenblatt transformations, \bt{which are evaluated via kernel density estimation of the associated conditional cumulative distributions. Furthermore, we} extended the construction to random field data via the Karhunen-Lo{\`e}ve expansion. We then take advantage of closed-form solutions for computing PCE Sobol indices to perform a global sensitivity analysis of the model which quantifies the intrinsic noise contribution to the overall model output variance. Additionally, the resulting joint PCE is generative in the sense that it allows generating random realizations at any input parameter setting that are statistically approximately equivalent to realizations from the underlying stochastic model. The method is demonstrated on a chemical catalysis example model \bt{and a synthetic example controlled by a parameter that enables a switch from unimodal to bimodal response distributions}.\\[5pt]

\end{abstract}


\section{Introduction}\label{sec:intro}

Engineering and applied science rely on computational simulations to rigorously study physical systems. These mathematical models are highly complex and their simulation parameters are often not known exactly. Usually, rigorous uncertainty quantification (UQ) studies are sample-intensive, i.e. they require many evaluations of the model at various conditions and parameter settings. The space of simulation conditions or parameters (`inputs') is sampled extensively and the model evaluated at each input to obtain a model response. The set of inputs and associated model responses is then probed, for example through analysis of variance methods, to explore how the input uncertainty influences the output. 

However, computational models of complex physical phenomena are often expensive and typically not amenable to intensive sampling schemes. In order to alleviate this expense, one often constructs a \emph{surrogate model} of the input-output map of the underlying physical model. The surrogate is constructed using a feasible number of training simulations spanning a range of model inputs. The surrogate model then provides a reasonably good approximation to the original physical model but the cost of model evaluation is comparatively negligible. Thus, the surrogate can be queried in place of the original model in order to perform sample-intensive studies necessary for UQ.

Popular surrogate choices that exist for deterministic models with uncertain inputs include kriging~\cite{Rasmussen:2005, Sack:1989}, radial basis functions~\cite{Gutmann:2001}, and support vector regression~\cite{Vapnik:1998, Smola:2004}. Polynomial chaos expansions (PCEs) have been particularly successful for surrogate modeling and forward/inverse UQ for a range of engineering applications~\cite{Ghanem:1991, OlmOmk:2010, Najm:2009a, Marzouk:2007, Marelli:2014,Debusschere:2016, Liang:2020, Robbe:2023b, Zhou:2024}. They serve as a convenient technique for uncertainty propagation, moment estimation, and variance decomposition due to the orthogonality property of polynomial bases~\cite{Sudret:2008,Crestaux:2009}.

We note that the outlined forward UQ workflow and the conventional surrogate options above are for \emph{deterministic} models with uncertain inputs, where only one model evaluation per input is necessary to fully characterize the model response. In contrast, in practical applications many examples arise where the model itself is inherently random. That is, uncertainty in the output is only partially explained by the uncertain input, and there is remaining uncertainty due to intrinsic noise that is not explicitly modeled. Multiple evaluations of such \emph{stochastic} models yield an uncertain response for any fixed input, and therefore the model response itself is a random variable. Forward UQ for such stochastic systems \bt{typically} requires repeated runs of the model for every computational condition in order to fully characterize the model response in distribution~\cite{Zhu:2020, Zhu:2021a}, \bt{although that restriction has been lifted in recent work~\cite{Zhu:2021b, Zhu:2023}}. \bt{Nevertheless, the intrinsic stochasticity} makes the computational burden of obtaining sufficient numbers of evaluations even more prohibitive. 

While local/wavelet-based PCE constructions can be used to represent non-smooth functions \cite{LeMaitre:2003a, Marelli:2021}, in most practical applications the effectiveness of (global) PCEs relies on the smoothness of the input-output map comprising the model. As such, they are not well suited for noisy/non-smooth input-output maps. As a remedy, PCEs constructed for noisy systems have typically been built for integrated output quantities of interest (QoIs), e.g.\ expectations of model output~\cite{Marrel:2012, Binois:2018} or quantiles of interest~\cite{Koenker:1978, Plumlee:2014, Torossian:2020}.  

In contrast, surrogates for stochastic models can be devised using a number of alternate approaches. For example, random processes can be optimally represented using the Karhunen-Lo{\`e}ve expansions (KLE)~\cite{Karhunen:1946,Ghanem:1991,LeMaitre:2010,Sargsyan:2010}, a generative construction that captures the first and second order two-point statistics of a random process. Alternatively, distribution-based constructions are available. For example, the Rosenblatt transformation can be used to provide a mapping between random vectors and associated standard random variables via conditional cumulative distribution functions (CDFs)~\cite{Rosenblatt:1952}. Further, optimal transport theory and transport maps have been used extensively to construct mappings between probability density functions (PDFs)~\cite{Marzouk:2016}. Analogously, methods which learn a conditional target density from data such as normalizing flows~\cite{Rezende:2015} have found use in machine learning.

It is observed in~\cite{Baker:2022} that recent surrogate constructions designed to predict the full response PDF, rather than only the output, often have difficulties producing uncertainties and identifying critical inputs. This limits their usefulness in a UQ context.

In this work, we generally strive to delay the averaging and consequent loss of information until after the surrogate is constructed. That is, we would like to develop a surrogate approximation to the stochastic model that can generate replica samples with requisite statistics/distributions. We further seek a surrogate model construction that provides a natural connection between the model inputs and these output realizations, while also allowing straightforward decomposition of uncertainties.

Broadly, there are two approaches for achieving this. The first utilizes a `multi-stage' design that captures the influence of the intrinsic noise on the QoI by extracting a \bt{deterministic} set of summary statistics for the distribution of noisy outputs. Methods described above are then used to directly map model inputs to these summary statistics~\cite{Baker:2022, Moutoussamy:2015, Zhu:2020, Mueller:2023}. Typically, a set of sample replica data is necessary to provide information about the noise distribution. The second approach more directly fits the model response to a distribution using either parametric or non-parametric strategies. 

\bt{Non-parametric methods such a kernel density estimation (KDE) may be used to represent arbitrary types of response distributions~\cite{Hall:2004} without imposing \textit{a priori} assumptions on the form of the PDF. However, non-parametric methods suffer from the curse of dimensionality, potentially requiring many sample replicas when representing stochastic models. Parametric methods are more restrictive, fitting the data to known distribution families, but usually require fewer samples. Notably,~\cite{Zhu:2020} employs the highly expressive parametrized class of generalized-lambda distributions, while~\cite{Zhu:2023} relies on PCEs as parameterized family of distributions. In fact, recent work removes the requirement of having multiple replicas for a given parameter value by employing moment estimation of generalized lambda distributions~\cite{Zhu:2021b} or a quadrature-based integration within a PCE construction~\cite{Zhu:2023} while implicitly relying on information present in the dataset across multiple parameter settings.} 

\bt{We observe that several of these approaches (see, e.g.~\cite{Moutoussamy:2015,Zhu:2020,Zhu:2021b}) utilize PCEs \emph{in conjunction} with other methods to construct a surrogate model, resulting in stochastic surrogates which are not \bt{necessarily} PCEs themselves. Building on the latest work~\cite{Zhu:2023,Luthen:2023,Mueller:2023}, however, we remain in the full PCE framework by employing the Rosenblatt transformation~\cite{Rosenblatt:1952} to incorporate the generative, stochastic dimensions while keeping the surrogate approximation valid in the parametric dimensions.}

A key element of UQ is assessing the sensitivity of the model output to the uncertainty of the input~\cite{Saltelli:2004}. While local approaches for sensitivity analysis exist, the validity of these methods over a range of values of uncertain variables is limited. Global sensitivity analysis (GSA) methods are required to accurately represent sensitivities when input uncertainties are large, the model is non-linear, and/or interactions exist between model inputs~\cite{Saltelli:2019}. Particularly, variance-based GSA methods exist, most commonly relying on Sobol sensitivity indices, which decompose the output variance and identify influential parameters by apportioning output variance to each uncertain input.

In most cases, Sobol indices cannot be found analytically and are estimated via sampling methods. However, global sensitivity indices computed from PCEs are analytically available and allow indices to be computed without sampling error~\cite{Sargsyan:2010, Sargsyan:2016, Sudret:2008, Nagel:2020}. By remaining in the PCE framework, we have immediate access to analytical expressions for global sensitivity indices and consequent variance decomposition into parametric uncertainty and intrinsic noise contributions.  That is, while developed in the deterministic input-output map framework, Sobol indices can be extended to stochastic systems in natural ways. As pointed out in~\cite{Zhu:2021a}, the treatment of the intrinsic noise alters the definition of the Sobol index. Typically, estimation of sensitivity indices is approached from one of three common viewpoints: (i) the intrinsic noise is represented with latent (hidden) variables that are included with the set of uncertain model inputs~\cite{Iooss:2009}; (ii) noise is eliminated by defining integrated QoIs and constructing the surrogate for these; or (iii) the indices are defined as functions of the latent (hidden) variables and are random quantities~\cite{Hart:2017, Jimenez:2017}. Sensitivity indices calculated for our PCE surrogate, constructed on a joint stochastic-parametric space, can be reformulated under each of these definitions and are useful for gauging the contribution of intrinsic noise to output variability.

The following text is organized as follows. In Section~\ref{sec:methods}, we describe the basic building blocks of the methodology, PCEs and Rosenblatt maps, and the joint PCE construction for vector QoIs. We then introduce the \bt{KLE} for random field quantities in Section~\ref{sec:field}, and briefly show how the joint PCE extends to field QoIs in a natural way. Next, Section~\ref{sec:results} describes the application problem of interest and results of our method applied to it, \bt{as well as a synthetic example with a complex stochastic nature}. We close with concluding remarks in Section~\ref{sec:concl}.

\section{Methodology}\label{sec:methods}

Consider a stochastic model $\bm{y} = f(\bm{\lambda}, \omega)$ where the forward function $f$ depends on a set of input parameters $\bm{\lambda} \in \mathbb{R}^{\tilde{d}}$ and outputs a random vector $\bm{y} \in \mathbb{R}^{d}$. With $\omega$ we denote an element of the sample space explicitly indicating the stochasticity of the model. The underlying assumption is that $f$ is expensive to evaluate directly, and therefore it is useful to identify a simpler model whose input-output behavior closely mimics the underlying forward function. The simple model acts as a surrogate representation for $f$ and can be used in its place to perform sample-intensive studies of the system of interest such as uncertainty propagation, sensitivity analysis, or parameter estimation.

\subsection{Polynomial Chaos Expansions} \label{sec:pce}
Polynomial chaos expansions are functional representations of random variables and provide a convenient framework for UQ in computational models~\cite{Ghanem:1991,LeMaitre:2001b,Debusschere:2003d,Xiu:2003b,Najm:2009a,Ghanem:2017}. Typically, uncertain inputs are written as a polynomial expansion \\[-0.315in]
\begin{align} \label{eq:input_pce}
\bm{\lambda}\simeq \sum_{p=0}^{P^{\prime}-1} \bm{a}_{p} \Psi_{p}(\bm{\xi})
\end{align}
in terms of multivariate polynomials $\Psi_{p}(\bm{\xi})$, where $\bm{\xi}=(\xi_1, \dots, \xi_{\tilde{d}})$ is a vector of independent identically distributed (\emph{i.i.d.})~standard random variables and the $\Psi_p(\bm{\xi})$ polynomials are orthogonal with respect to the density of $\bm{\xi}$. The multivariate polynomials in turn are written as products of univariate polynomials $\Psi_{p}(\bm{\xi}) = \prod_{\ell=1}^{\tilde{d}} \psi_{p_{\ell}}(\xi_{\ell})$, where each scalar index $p=p(p_1, \dots, p_{\tilde{d}})$  accounts for the corresponding \emph{multi-index} or multivariate degree $(p_1, \dots, p_{\tilde{d}})$. 

\bt{While construction of customized PCE surrogates can be quite efficient~\cite{Witteveen:2006a, Oladyshkin:2012, Torre:2019}, we employ Legendre-Uniform PCEs (Legendre polynomials as functions of standard uniform random variables), since we} are targeting uniform point accuracy of the surrogate. Moreover, we assume a first-order, independent-component PCE for each parameter $\lambda_i= \frac{b_{i} + a_{i}}{2} + \frac{b_{i} - a_{i}}{2} \xi_{i}$ given a physically meaningful interval of uncertainty $[a_i, b_i]$, for $i=1, \dots, \tilde{d}$. This simplifying assumption is made without loss of generality: we largely rely on a correspondence $\bm{\lambda}=\bm{\lambda}(\bm{\xi})$ between the physical parameter $\bm{\lambda}$ and the parametric \emph{germ} $\bm{\xi} \in \text{Unif}[-1,1]^{\tilde{d}}$.

Conventionally, for a \emph{deterministic} model $g(\bm{\lambda})$, one writes an associated PCE representation \\[-0.2in]
\begin{align} \label{eq:output_pce}
g(\bm{\lambda}) \simeq \sum_{p=0}^{P-1} \bm{c}_{p} \Psi_{p}(\bm{\xi}) \\[-0.3in] \nonumber
\end{align}
with the same germ $\bm{\xi}$, and finds coefficients $\bm{c}_{p}$ by regression using a set of $N$ model evaluations \bt{$g(\bm{\lambda}^{(n)})$} at select training inputs \bt{$\bm{\lambda}^{(n)}=\bm{\lambda}(\bm{\xi}^{(n)})$}, for $n=1, \dots, N$. Together with the map $\bm{\lambda}(\bm{\xi})$, the PCE in Eq.~\eqref{eq:output_pce} acts as a surrogate representation for the deterministic model $\bm{y} = g(\bm{\lambda})$~\cite{Sargsyan:2016}. 

Representing random variables with PCEs offers a number of advantages. Due to the orthogonality of the basis functions, one can show that the mean of the random variable is the zeroth-order coefficient, and its variance is a weighted sum-of-squares of the rest of the coefficients:\\[-0.2in]
\begin{align} \label{eq:pce_moments}
\mathbb{E}[\bm{y}] = c_{0}, \qquad \mathbb{V}[\bm{y}] = \sum_{p=1}^{P-1} c_{p}^{2} || \Psi_{p} ||^{2}, \\[-0.25in] \nonumber
\end{align}
where the norm $\|\Psi_p\|$ is defined as weighted with the density of $\bm{\xi}$, i.e.
\begin{align} \label{eq:normed_psi}
\|\Psi_p\|^2:=\int \Psi_p(\bm{\xi})^2 \rho_{\bm{\xi}}(\bm{\xi})\mathrm{d}\bm{\xi}.
\end{align}

Furthermore, PCE-based surrogate models enable variance-based decomposition and subsequent extraction of Sobol indices. The Sobol indices decompose the output variance into fractional variance contributions of each parameter or group of parameters~\cite{Saltelli:2006}. Here we examine the main-effect (first-order) sensitivities, which measure the influence of each parameter in isolation on the output variance. These indices are defined as variance ratios and can be computed exactly from the PCE coefficients:
\begin{align} \label{eq:main_sens}
S_{i} &:= \frac{\mathbb{V}_{\lambda_{i}} [\mathbb{E}_{\lambda_{-i}}[ g(\bm{\lambda} \, | \, \lambda_{i})]]}{\mathbb{V}[\bm{y}]} 
	= \frac{1}{\mathbb{V}[\bm{y}]} \sum_{p \in \mathcal{P}_{S_{i}}} c_{p}^{2} \vert \vert \Psi_{p} \vert \vert ^{2}, \\
	&\text{where  } \mathcal{P}_{S_{i}} := \{p : p_{i}(p) > 0,\  p_{j}(p) = 0 \text{  for  } j \neq i \} \nonumber
\end{align}
where $\mathbb{V}_{\lambda_{i}}$ denotes the variance with respect to the $i$th parameter and $\mathbb{E}_{\lambda_{-i}}$ is the expectation with respect to all remaining parameters. Similar to the first two moments in Eq.~\eqref{eq:pce_moments}, the right-hand-side of Eq.~\eqref{eq:main_sens} is derived using the orthogonality of PCE bases~\cite{Crestaux:2009, Sargsyan:2016, Sudret:2008}. 

\subsection{The Rosenblatt Map and Stochastic PCEs}

 Underpinning the PCE surrogate construction for the deterministic forward model described above is the explicit mapping between uncertain parameters $\bm{\lambda}$ -- and consequently, model outputs $g(\bm{\lambda})$ -- and the PCE germ $\bm{\xi}$. In this work, we refer to these PCE surrogates as `parametric PCEs.'

Now, suppose that our model has intrinsic noise, i.e. for each \emph{fixed} value of $\bm{\lambda}$ its output is a random quantity $f(\bm{\lambda})$. For ease of presentation, we write the forward function as $f(\bm{\lambda}, \omega)$ to explicitly highlight a sample space element $\omega$.  Surrogate construction is not as straightforward as described in Section~\ref{sec:pce}, since there is no explicit control on $\omega$. Having said that, one can construct a PCE representation of a random variable $\bm{y}=f(\bm{\lambda}, \omega)\in \mathbb{R}^{d}$ given $M$ samples $\{\bm{y}^{(m)}=f(\bm{\lambda}, \omega^{(m)})\}_{m=1}^M$ relying on the Rosenblatt transformation~\cite{Rosenblatt:1952} and \bt{KDE}~\cite{Silverman:1986} \bt{to construct a mapping through the conditional CDF $\mathcal{F}$}. The Rosenblatt transformation is the multivariate generalization of the cumulative distribution function theorem, 
\begin{align} 
\nonumber
\zeta_{1} &= \mathcal{F}_{1}(y_{1}) \\
\nonumber
\zeta_{2} &= \mathcal{F}_{2|1}(y_{2} | y_{1}) \\
\zeta_{3} &= \mathcal{F}_{3|2,1}(y_{3} | y_{2}, y_{1}) \\[-2.5mm]
\nonumber
&\hspace{1.75mm} \vdots \\[-2.5mm]
\nonumber
\zeta_{d} &= \mathcal{F}_{d|d-1,\dots,2,1}(y_{d} | y_{d-1},\dots, y_{2}, y_{1}),
\end{align}
mapping any $d$-dimensional random vector $\bm{y}$ to the $d$-dimensional random vector $\bm{\zeta}$ composed of \emph{i.i.d.}\ uniform random variables $\bt{\bm{\zeta}=}(\zeta_1, \dots, \zeta_d)$. In fact, the inverse of the Rosenblatt transformation $R_{\bm{\lambda}}^{-1}(\bm{\zeta})=\bm{y}$ essentially provides a mapping from $\omega$ to the germ $\bm{\zeta}$ (see~\cite{Sargsyan:2010} for details). \bt{While one could associate a different germ $\bm{\zeta}(\vlam)$ with each $\vlam$}, \bt{h}ere we make the modeling choice of representing the possible $\bm{\zeta}(\vlam)$ with a single underlying germ \bt{vector} $\bm{\zeta}$ that has stochastic dimension $d$, \bt{i.e.~the same dimensionality as $\bm{\lambda}$}. Essentially, this pushes all dependence on $\vlam$ into the expansion coefficients which we denote $\vz_{s}(\vlam)$. In other words, $R_{\bm{\lambda}}^{-1}$ acts as a forward map allowing construction of the PCE similar to Section~\ref{sec:pce}. Consequently, the PCE is written, for any $\bm{\lambda}$, as \\[-0.25in]
\begin{align} \label{eq:stochastic_pce}
\bm{y} =  {R}_{\bm{\lambda}}^{-1}(\bm{\zeta}) \simeq \sum_{s=0}^{S-1} \bm{z}_{s}(\vlam) \Psi_{s}(\bm{\zeta}).
\end{align}
\bt{In principle the total truncation order is $\bm{\lambda}$-dependent, but in this work we assume the same order for all $\bm{\lambda}$ for simplicity of presentation. The constant $S$ is the number of terms included in each expansion and is the result of the selected total truncation order which can be chosen empirically.} This expansion is built purely on samples of the intrinsic noise for a fixed parameter value $\vlam$ and, consequently, we refer to it as a `stochastic PCE.' 

\bt{This stochastic PCE is constructed for each $\vlam$ via least-squares regression. The Rosenblatt transformation, evaluated by KDE of the associated CDFs, enables obtaining training pairs $\{\bm{\zeta}, R^{-1}_{\vlam}(\bm{\zeta})\}$ for the regression. The underlying distributions (i.e. CDFs) of intrinsic noise are explicitly represented, informed by sample replica data. All dependence on the parameters $\bm{\lambda}$ is collected in the stochastic expansion coefficients $\bm{z}_{s}(\bm{\lambda})$ by choice of the underlying germ $\bm{\zeta}$. These $\bm{z}_{s}(\bm{\lambda})$ act as deterministic summarizing quantities for the distribution (similar to generalized lambda distribution parameters in~\cite{Zhu:2020}, yet still possessing the generality of PCEs), enabling the construction of a second, pointwise mapping $\bm{\lambda} \mapsto \bm{z}_{s}(\bm{\lambda})$ (see Section \ref{sec:joint_pce}).} 

\subsection{The joint PCE Surrogate} \label{sec:joint_pce}

With the parametric and stochastic PCE constructions in mind, we now describe the algorithm (see also \cite{Mueller:2023}) to arrive at a PCE surrogate on a joint space $(\bm{\lambda}, \omega)$ of parametric and stochastic dimensions. 

Suppose we have an ensemble of parameter configurations $\bm{\lambda}^{(n)} \, (n=1,\dots,N)$ and, for each fixed $\bm{\lambda}^{(n)}$, we obtain $M$ samples of the stochastic model $\{\bm{y}^{(n,m)}=f(\bm{\lambda}^{(n)}, \omega^{(n,m)})\}_{m=1}^M$ where $n$ is fixed. For each parametric sample, we build a multi-output stochastic PCE according to Eq.~\eqref{eq:stochastic_pce} which takes the form \\[-0.15in]
\begin{align} \label{eq:stochastic_pce2}
f(\bm{\lambda}^{(n)}, \omega) \simeq \sum_{s=0}^{S-1} \bm{z}_{s}(\vlam^{(n)}) \Psi_{s}(\bm{\zeta}), \quad n =1,\dots,N.
\end{align}

The next step in our joint PCE construction is to build a parametric PCE for the coefficients $\bm{z}_{s}(\vlam)$, $s=0,\dots, S-1$, where the samples $\{\vlam^{(n)}, \bm{z}_{s}(\vlam^{(n)})\}_{n=1}^N$ will serve as training data for the parametric PCE regression. \bt{One can envision adaptive approaches for identifying optimal orders for each $s$} \cite{Zhu:2021b,Zhu:2023}\bt{; in this work we rely on model evidence which is analytically available as the model is linear with respect to coefficients of the PCE~\cite{Sargsyan:2016}, providing a rigorous way of finding the optimal truncation order, and consequently, the number of terms $P$}. 
The PCE then takes the form
\begin{align} \label{eq:coefficient_pce}
\bm{z}_{s}(\bm{\lambda}) \approx \bm{z}_{s}^{\text{\tiny PCE}}(\bm{\lambda}) \simeq \sum_{p=0}^{P-1} \bm{a}_{sp} \Psi_{p}(\bm{\xi}),
\end{align}
\bt{where the coefficients $a_{sp}$ are found via least-squares regression.} Replacing the coefficients in Eq.~\eqref{eq:stochastic_pce2} with this coefficient surrogate Eq.~\eqref{eq:coefficient_pce}, we obtain a joint stochastic-parametric PCE representation
\begin{align} \label{eq:sequential_pce}
f(\bm{\lambda},\omega) \simeq \sum_{s=0}^{S-1} \bigg( \sum_{p=0}^{P-1} \bm{a}_{sp} \Psi_{p}(\bm{\xi}) \bigg)\Psi_{s} (\bm{\zeta}).
\end{align}
In other words, we arrive at a joint PCE by building regression-based stochastic PCEs from inverse-Rosenblatt maps for each parameter $\bm{\lambda}$ followed by another regression with respect to the parameter $\bm{\lambda}$ itself. Note that one can rewrite the nested expansion Eq.~\eqref{eq:sequential_pce} as a single joint expansion with a higher-dimensional multivariate basis 
\begin{align} \label{eq:joint_pce}
f(\bm{\lambda},\omega) \simeq \sum_{j=0}^{J-1} \bm{c}_{j} \Psi_{j}(\bm{\xi}, \bm{\zeta})
\end{align}
where the PCE germs $\bm{\xi}$ and $\bm{\zeta}$ correspond to their `physical' counterparts $\bm{\lambda}$ and $\omega$, respectively. \bt{The single sum~\eqref{eq:sequential_pce} and double sum~\eqref{eq:joint_pce} representations of the surrogate model are equivalent up to the choice of the multi-indices. Note that the joint PCE Eq.~\eqref{eq:joint_pce} is equivalent to that of Eq.~(13) of \cite{Zhu:2023}. The construction of the joint surrogate in~\cite{Zhu:2023} proceeds via direct maximization of a likelihood function relying on the introduction of data noise, without need for sample replicas, and with associated integration over conditional probabilities via quadrature at each parametric input.} 

\bt{We observe that within our framework \eqref{eq:joint_pce} can be directly achieved, too, by first associating $\omega$ with a given germ $\bm{\zeta}$ through the Rosenblatt transformation, as before, and then directly constructing the stochastic-parametric surrogate via regression in the joint space of $(\bm{\xi},\bm{\zeta})$ parameters. The eventual surrogate obtained via least-squares regression in the joint space would produce equivalent results up to a truncation rule of the multi-indices. Nevertheless, we opt for the sequential construction due to the extra flexibility allowed by having a separate truncation rule for each parameter, even though it comes with an additional requirement of having replications per parameter sample in order to evaluate the inverse Rosenblatt transform accurately.} 

The mixed PCE in Eq.~\eqref{eq:joint_pce} offers a number of advantages. It can be viewed as a generative model while remaining statistically consistent with the underlying model for each parameter value. Besides, it allows additional flexibility of employing mixed bases. That is, while bounded parameter ranges naturally lead to \bt{Legendre-Uniform} bases in the $\bm{\xi}$ dimensions, one can employ Gauss-Hermite bases in the $\bm{\zeta}$-dimensions corresponding to intrinsic stochasticity as the latter is often expected to have Gaussian behavior for many physical problems of interest due to aggregation of multiple independent contributions to intrinsic noise. 

Additionally, the joint PCE Eq.~\eqref{eq:joint_pce} is flexible enough to allow various interpretations of GSA for stochastic models. Namely, as categorized in~\cite{Zhu:2021a}, the variance-based GSA for stochastic models can be approached from one of three viewpoints: (i) the intrinsic noise is summarized with a (set of) latent variable(s) that are included with model inputs and the Sobol indices are functions of these variables \cite{Iooss:2009}; (ii) the intrinsic noise is first eliminated by averaging processes which represent the model response with one or more summary statistics (QoIs) for the system and the Sobol indices are functions of the QoIs; or (iii) the intrinsic noise is associated with (deterministic) realizations of the model obtained by holding the random seed fixed, and obtaining Sobol indices as functions of the random seed \cite{Hart:2017}.  

Our treatment of the intrinsic noise and resulting Sobol indices most closely aligns with viewpoint (i), although it is flexible enough to be framed by viewpoints (ii) and (iii) also. 
That is, framing the intrinsic noise as a random variable $\omega$ which we augment to the inputs $(\vlam,\omega)$ is analogous to viewpoint (i). We can extract variance-based global sensitivity indices using Eq.~\eqref{eq:main_sens} and attribute the variance in the output to either uncertainties in the parameters or intrinsic noise from the stochastic model. Specifically, the Sobol index from the intrinsic model noise is essentially a \emph{group sensitivity} that is found by summing the effects of all $\bm{\zeta}$.

Furthermore, the QoI-based framework (ii) applies also if we treat the Rosenblatt coefficients from the stochastic PCE construction as summary QoIs. For example, in the first-order case, the Rosenblatt-PCE coefficients are identically the mean and the standard deviation, which are summary statistics for the system. The parametric PCE is constructed for these QoIs, effectively averaging out the intrinsic noise.

Finally, if we rearrange the sums in Eq.~\eqref{eq:sequential_pce} \\[-0.2in]
\begin{align*}
f(\bm{\lambda},\omega) \simeq \sum_{p=0}^{P-1} \bigg(\sum_{s=0}^{S-1} \bm{a}_{sp} \Psi_{s} (\bm{\zeta})\bigg) \Psi_{p}(\bm{\xi})
\end{align*}
we obtain parametric Sobol indices that are dependent on $\bm{\zeta}$, which is exactly the form of the indices obtained in viewpoint (iii). 

\section{Representing Random Fields}\label{sec:field}

So far, the stochastic models we have presented are defined over a joint space $(\vlam, \omega)$ of parametric and stochastic dimensions, and the \emph{deterministic coordinates} $\vx$ of the system -- such as the spatial domain and time evolution -- are fixed. This gives a single `snapshot' of the system at a given location and an instant in the model's evolution. To study the system progression across a continuous range of deterministic coordinates, we now consider stochastic processes $\textbf{Y} = f(\vlam, \omega; \vx)$ where the forward function $f$ depends explicitly on the system deterministic coordinates $\vx \in \mathbb{R}^{L_{\bm{x}}}$ and outputs a random field QoI $\textbf{Y} \in \mathbb{R}^{d \times L_{\bm{x}} }$. With $\vlam$ and $\omega$ we denote the input parameters and model stochasticity, as before.  The dimension $L_{\bm{x}}$ encapsulates any combination of spatial and/or temporal dimensions such as $N_{\bm{x}} \times N_{\bm{y}}$ or $N_{\bm{x}} \times N_{\bm{t}}$ or $(N_{\bm{x}} \times N_{\bm{y}}) \times N_{\bm{t}}$.

We rely on \bt{KLE} to arrive at a compact representation of the random field~\cite{Karhunen:1946,Ghanem:1991,Ghanem:1999c,Huang:2001,Sargsyan:2010}. The algorithm described in Section \ref{sec:methods} is then easily applied to the expansion leading to a joint PCE surrogate that can be queried as before for sample-intensive UQ studies of the system. 

\subsection{Compact representation of fields with KLE}

The Karhunen-Lo{\`e}ve expansion is a well-established spectral expansion technique that expresses a random field as weighted sums of zero-mean orthonormal random variables, \\[-0.225in]
\begin{align}\label{eq:inf_kle}
	f(\vlam, \omega; \vx) &= f_{0}(\vx) + \sum_{l=1}^{L_{\bm{x}}} \eta_{l}(\vlam,\omega) \sqrt{\mu_{l}} \, \phi_{l}(\vx), \qquad L_{\bm{x}}< \infty.
\end{align}
Here, the expectation $f_{0}(\vx) := \mathbb{E}_{\lambda, \omega}[f(\vlam, \omega; \vx)]$ centers the process. The $(\mu_{l}, \phi_{l}(\vx))$ are (deterministic) eigenvalue-eigenfunction pairs of the sample covariance 
\begin{align*}
	\mathbb{C}(\vx,\vx^{\prime}) = \mathbb{E}_{\lambda,\omega}[(f(\vlam, \omega; \vx) - f_{0}(\vx))(f(\vlam, \omega; \vx^{\prime}) - f_{0}(\vx^{\prime}))]
\end{align*}
and the eigenvalues are ordered $\mu_{1} \geq \mu_{2} \geq \dots \geq 0$. The random coefficients $\eta_{l}(\vlam,\omega)$ are uncorrelated with zero mean and unit variance~\cite{Karhunen:1946, Ghanem:1991, Ghanem:1999c, Huang:2001, Sargsyan:2010, Luthen:2023}. 

While the expansion is infinite by definition, in a discretized $\bm{x}$ context it becomes exact once the number of included eigenterms $L_{\bm{x}}$ equals the number of deterministic coordinates, or the length of the discretized $\bm{x}$ vector.  In practice, the sum in Eq.~\eqref{eq:inf_kle} is truncated to obtain a lower-dimensional representation of the random field \\[-0.275in]
\begin{align}\label{eq:kle}
	f(\vlam, \omega; \vx) &\approx f_{0}(\vx) + \sum_{l=1}^{L} \eta_{l}(\vlam,\omega) \sqrt{\mu_{l}} \, \phi_{l}(\vx), \quad L \ll L_{\bm{x}}.
\end{align}

Identification of a truncation point $L$ exploits simple expressions for the model variance. First, the pointwise model variance is computed from the expansion at each $\vx$ by taking advantage of independence of $\eta_{l}$ and orthonormality of $\phi_{l}(\vx)$, \\[-0.3in]
\begin{align}
\mathbb{V}_{\bm{\lambda}, \omega}[f(\vx)] = \sum_{l=1}^{L_{\bm{x}}}  \mu_{l} \phi_{l}(\vx)^{2}_{\bm{.}} 
\end{align}
Integrating the pointwise variance over the set of deterministic coordinates $\vx \in \Omega$ gives the total output variance as the sum of the eigenvalues, i.e. \\[-0.225in]
\begin{align}
\mathbb{V}_{\bm{x}}[f]  
	= \int_{\Omega} \sum_{l=1}^{L_{\bm{x}}} \mu_{l} \phi_{l}(\vx)^{2} d\vx 
	= \sum_{l=1}^{L_{\bm{x}}}  \mu_{l} .
\end{align}

We utilize the natural ordering of the eigenvalues to obtain the fraction of the total output variance explained by the first $L$ terms of the expansion
\begin{align} \label{eq:fracvar}
	\frac{\sum_{l=1}^{L} \mu_{l}}{\sum_{l=1}^{L_{\bm{x}}} \mu_{l}} = 1 - \varepsilon, \quad 0 < \varepsilon < 1.
\end{align}
From this, the truncation point is chosen so that $\varepsilon$ is negligibly small, for example $\varepsilon < 0.01$ or at least $99\%$ of the explained variance is preserved by the KLE. 

Note that the number of terms needed in the KLE depends intimately on the rate of decay of the eigenvalues. If the field is smooth, then the spectrum will rapidly decay and only the first few expansion terms are needed to represent the random field with reasonable accuracy. On the other hand, spectral decay is slow and a large number of expansion terms are needed to represent rough fields (e.g. Gaussian white noise). 

\subsection{Joint PCE surrogates for random fields}

\bt{There is a large body of literature extending the PCE surrogate construction to vector QoIs or random field QoIs $\textbf{Y}=f(\vlam, \omega; \vx) \in \mathbb{R}^{d \times L_{\bm{x}}}$ \cite{Marzouk:2009, Das:2009, Sargsyan:2010, Blatman:2013, Betz:2014, Azzi:2019, Luthen:2023}. Here we build on these approaches in the context of the developed joint stochastic-parametric PCE.} Suppose we have a collection of parameter configurations $\vlam^{(n)}, n=1,\dots,N$, and for each fixed $\vlam^{(n)}$ we obtain $M$ samples of the random field $f(\vlam^{(n)}, \omega^{(n,m)}; \vx)$. From this, we wish to construct a joint PCE surrogate in the form of Eq.~\eqref{eq:joint_pce} using the methods described in Section~\ref{sec:methods}.

We first find a low dimensional representation via the KLE for the random field QoI(s) using Eq.~\eqref{eq:kle}. The expansion is \emph{global} in the sense that the set of \emph{combined} stochastic-parametric samples are taken together in this expansion. The expansion coefficients $\eta_{l}(\vlam,\omega)$ are random variables and will be represented with joint PCE surrogates: \\[-0.225in]
\begin{align}\label{eq:jpce_eta}
	\eta_{l}(\vlam,\omega) \simeq \sum_{j=0}^{J-1} \bm{c}_{lj} \Psi_{j}(\bm{\xi},\bm{\zeta}), \quad \bt{\text{ for  } l = 1,...,L}. 
\end{align}

Substituting this joint PCE for $\eta_{l}(\vlam,\omega)$ into the KLE yields an expression for the random field QoIs.\\[-0.15in]
\begin{align}\label{eq:klpc}
	f(\vlam, \omega; \vx) &= f_{0}(\vx) + \sum_{l=1}^{L} \bigg(\underbrace{\sum_{j=0}^{J-1} \bm{c}_{lj} \Psi_{j}(\vxi, \vzeta)}_{\eta_{l}(\vlam, \omega)}\bigg) \sqrt{\mu_{l}} \, \phi_{l}(\vx) \\[-0.3in] \nonumber
\end{align}
Simple rearrangement of terms allows us to identify the surrogate model Eq.~\eqref{eq:klpc} as a PCE with coefficients that depend on the deterministic coordinates $\bm{x}$. We refer to the result as a \emph{KLPC surrogate} (i.e. a surrogate constructed via KLE and PCEs),
\begin{align}\label{eq:klpc_2}
	f(\vlam, \omega; \vx) 
	&= \sum_{j=0}^{J-1} \bigg(\underbrace{ f_{0}(\vx) \delta_{j0} + \sum_{l=1}^{L} \bm{c}_{lj} \sqrt{\mu_{l}} \, \phi_{l}(\vx) }_{ \vc^{\text{\tiny KLPC}}_{j}(\vx) } \bigg) \Psi_{j}(\vxi, \vzeta). \\[-0.25in] \nonumber
\end{align}
The Kronecker delta function $\delta_{j0}$ includes the process mean $f_{0}(\vx)$ in only the constant $(j=0)$ term of the KLPC.

\bt{Note that~\cite{Luthen:2023} proposes an alternate approach for constructing stochastic surrogate representations of random fields (i.e.~solution `trajectories'). However, while we first compress the random fields with KLE and then represent the result as PCEs,~\cite{Luthen:2023} employs a reverse mechanism by first representing discrete random field samples as PCEs and then capturing the underlying stochasticity of the model with KLE.}

Since the above KLPC is indeed a PCE, we take advantage of the analytical expressions for the moments and Sobol indices in terms of the PCE coefficients to perform GSA without incurring sampling error. \bt{Relying on the corresponding PCE-based moment calculations in Eq.~\eqref{eq:pce_moments}, the first and second moments of the KLPC given in Eq.~\eqref{eq:klpc_2} are}  \\[-0.25in]
\begin{align} \label{eq:klpc_moments}
	\mathbb{E}[\vy(\vx)] 
		&= c^{\text{\tiny KLPC}}_{0}(\vx) 
		= f_{0}(\vx) + \sum_{l=1}^{L} \bm{c}_{l0} \sqrt{\mu_{l}} \, \phi_{l}(\vx) \\
	\mathbb{V}[\vy(\vx)] 
		&= \sum_{j=1}^{J-1} \big(c^{\text{\tiny KLPC}}_{j}(\vx) \big)^{2} \vert\vert \Psi_{j}\vert\vert^{2}
		= \sum_{j=1}^{J-1} \bigg(\sum_{l=1}^{L} \bm{c}_{lj} \sqrt{\mu_{l}} \, \phi_{l}(\vx) \bigg)^{2} \vert\vert \Psi_{j}\vert\vert^{2}. 
\end{align}
Unlike the moment equations given in Eq.~\eqref{eq:pce_moments}, the moments for the KLPC now depend on a particular deterministic coordinate of the system $\vx$ via the KLPC coefficients $\bm{c}_{j}^{\text{\tiny KLPC}}(\vx)$.

\bt{Similarly,} main effect sensitivity indices are found from Eq.~\eqref{eq:main_sens} by substituting in the KLPC coefficients \bt{into PC-based sensitivity formulae~\eqref{eq:main_sens}, as also outlined in} \cite{Marelli:2015, Nagel:2020}. 
\begin{align} 
S_{i} 	
	&= \frac{1}{\mathbb{V}[\vy(\vx)]]} \sum_{j \in \mathcal{J}_{S_{i}}} \big(c^{\text{\tiny KLPC}}_{j}(\vx)\big) ^{2}  \vert \vert \Psi_{j} \vert \vert ^{2}
	= \frac{1}{\mathbb{V}[\vy(\vx)]} \sum_{j \in \mathcal{J}_{S_{i}}} \bigg(\sum_{l=1}^{L} \bm{c}_{lj} \sqrt{\mu_{l}} \, \phi_{l}(\vx)  \bigg)^{2}  \vert \vert \Psi_{j} \vert \vert ^{2}, \\ 
	&\text{where   }  \mathcal{J}_{S_{i}} := \{j : p_{i}(p) > 0,  p_{k}(p) = 0 \text{  for  } k \neq i \} . \nonumber
\end{align}
In the expressions for both the output variance and Sobol indices, the squared-norm $\vert \vert \Psi_{j} \vert \vert^{2}$ is defined analogously to Eq.~\eqref{eq:normed_psi}.

\section{Results}\label{sec:results}

To illustrate the methodology, we consider a heterogeneous gas-surface chemical catalysis system, which we model using a mesoscopic lattice kinetic Monte Carlo (KMC) construction~\cite{Andersen:2019,Pineda:2022}. We presume an elementary-step chemical kinetic mechanism employing a number of reactions, with uncertain rate constants. At the same time, KMC employs intrinsic noise that models the randomness of chemical reaction at the microscale. 

As an example problem, we consider CO oxidation on a RuO$_{2}$(110) surface, represented at atomic scales as a lattice with alternating coordinatively unsaturated ({\verb#cus#}) sites and bridge ({\verb#br#}) binding sites where unimolecular adsorption (desorption) of CO and dissociative adsorption (desorption) of O$_{2}$ may occur. The adsorbed species can make diffusive hops to neighboring vacant sites and nearest-neighbor O and CO can react, producing  gaseous CO$_{2}$. The system is modeled using 22 permissible processes, where a `process' on the lattice is specified by event type, chemical species, and lattice site(s) involved -- for example, diffusion of CO from a {\verb#cus#} site to a {\verb#br#} site (see \cite{Andersen:2019, Dopking:2017} for a more thorough discussion of the mechanism). 

\begin{table}[ht!]
\centering
\begin{tabular}{llll}
\multicolumn{2}{c}{Forward Processes} & \multicolumn{2}{c}{Reverse Processe} \\ \hline \hline
$[1]$ \ Adsorption: & CO $\xrightarrow[]{k_1}$ CO({\verb#cus#})   				& Desorption: 	& CO({\verb#cus#}) $\xrightarrow[]{k_{-1}}$ CO 				 \\
$[2]$ \ Adsorption: & CO $\xrightarrow[]{k_2}$ CO({\verb#br#}) 					& Desorption: 		& CO({\verb#br#}) $\xrightarrow[]{k_{-2}}$ CO				 \\
$[3]$ \ Adsorption: & O$_2$ $\xrightarrow[]{k_3}$ O({\verb#cus#}) + O({\verb#cus#})	& Desorption: & O({\verb#cus#}) + O({\verb#cus#}) $\xrightarrow[]{k_{-3}}$ O$_2$ 	 \\
$[4]$ \ Adsorption: & O$_2$ $\xrightarrow[]{k_4}$ O({\verb#br#}) + O({\verb#br#}) 	& Desorption: & O({\verb#br#}) + O({\verb#br#}) $\xrightarrow[]{k_{-4}}$ O$_2$		 \\
$[5]$ \ Adsorption: & O$_2$ $\xrightarrow[]{k_5}$ O({\verb#br#}) + O({\verb#cus#})   	& Desorption: & O({\verb#br#}) + O({\verb#cus#}) $\xrightarrow[]{k_{-5}}$ O$_2$	 \\
$[6]$ \ Diffusion:    & CO({\verb#cus#}) $\xrightarrow[]{k_6}$ CO({\verb#cus#})	   	&		& \\
$[7]$ \ Diffusion:    & CO({\verb#br#}) $\xrightarrow[]{k_7}$ CO({\verb#br#})	   	& 		& \\
$[8]$ \ Diffusion:    & CO({\verb#cus#}) $\xrightarrow[]{k_8}$ CO({\verb#br#})	   	& Diffusion:    & CO({\verb#br#}) $\xrightarrow[]{k_{-8}}$ CO({\verb#cus#}) 		 \\
$[9]$ \ Diffusion:    & O({\verb#cus#}) $\xrightarrow[]{k_9}$ O({\verb#cus#})	   	&		& \\
$[10]$ Diffusion:    & O({\verb#br#}) $\xrightarrow[]{k_{10}}$ O({\verb#br#})	   	&		& \\
$[11]$ Diffusion:    & O({\verb#cus#}) $\xrightarrow[]{k_{11}}$ O({\verb#br#})	   	& Diffusion:    & O({\verb#br#}) $\xrightarrow[]{k_{-11}}$ O({\verb#cus#}) 			 \\
$[12]$ Formation: & CO({\verb#cus#}) + O({\verb#cus#}) $\xrightarrow[]{k_{12}}$ CO$_2$  	&	& \\
$[13]$ Formation: & CO({\verb#br#}) + O({\verb#br#}) $\xrightarrow[]{k_{13}}$ CO$_2$ 		&	& \\
$[14]$ Formation: & CO({\verb#br#}) + O({\verb#cus#}) $\xrightarrow[]{k_{14}}$ CO$_2$ 		&	& \\
$[15]$ Formation: & CO({\verb#cus#}) + O({\verb#br#}) $\xrightarrow[]{k_{15}}$ CO$_2$ 		&	& \\ 
\hline \hline
\end{tabular}
\caption{Summary of processes involved in the oxidation of CO to CO$_2$ via a RuO$_2$ catalyst. Forward reactions are given in the left column, and any associated reverse reactions listed in the right column. Adsorbed species are indicated by the type of binding sites: \texttt{br} or \texttt{cus}.  The rates of the forward (reverse) reactions are denoted above the reaction arrows by $k_{i}$ ($k_{-i}$). While diffusion is not typically written as a reaction in this way, we include the diffusion events for completeness of the mechanism. Here the reaction arrows associated with diffusion events indicate change of position on the lattice and not change to chemical structures.}
\label{tab:rxns}
\end{table}

Associated with each process is a reaction rate constant which is treated as an uncertain parameter. The CO$_{2}$ production reactions, as well as the species diffusion processes between identical lattice sites, are modeled as irreversible `forward' processes with associated forward rate constants $k_{f}$, while diffusion between non-identical lattice sites and adsorption/desorption are treated as reversible processes. The reversible processes have forward and reverse rate constants $k_{f}$ and $k_{r}$ that are thermodynamically balanced. The set of included reactions are summarized in Table \ref{tab:rxns}.

The uncertain input parameter vector $\bm{\lambda}$ consists of the logarithms of the $15$ forward reaction rates, specified as uniform random variables around given nominal values with a scale factor vector $\bm{r}$:
$$
\bm{\lambda} := \{ \bm{k} \, \vert \, \ln \bm{k} \sim \text{Unif}(\ln \bm{k}_{\text{\tiny nom}} - \ln \bm{r}, \ln \bm{k}_{\text{\tiny nom}} + \ln \bm{r})\}.
$$

The QoI considered here is the production rate of CO$_2$ over time which we define as $y(t) = n_{p}/(\alpha \Delta t)$ where $n_{p}/\Delta t$ is the number of CO$_2$ formations per time increment and $\alpha$ is the lattice area. That is, the production rate is given by the number of CO$_2$ formation events per time per unit area. Obviously, $\alpha$ will scale with the number of lattice cells $n_{c}$.

\begin{figure}[t!] 
\centering
\includegraphics[width=0.75\textwidth]{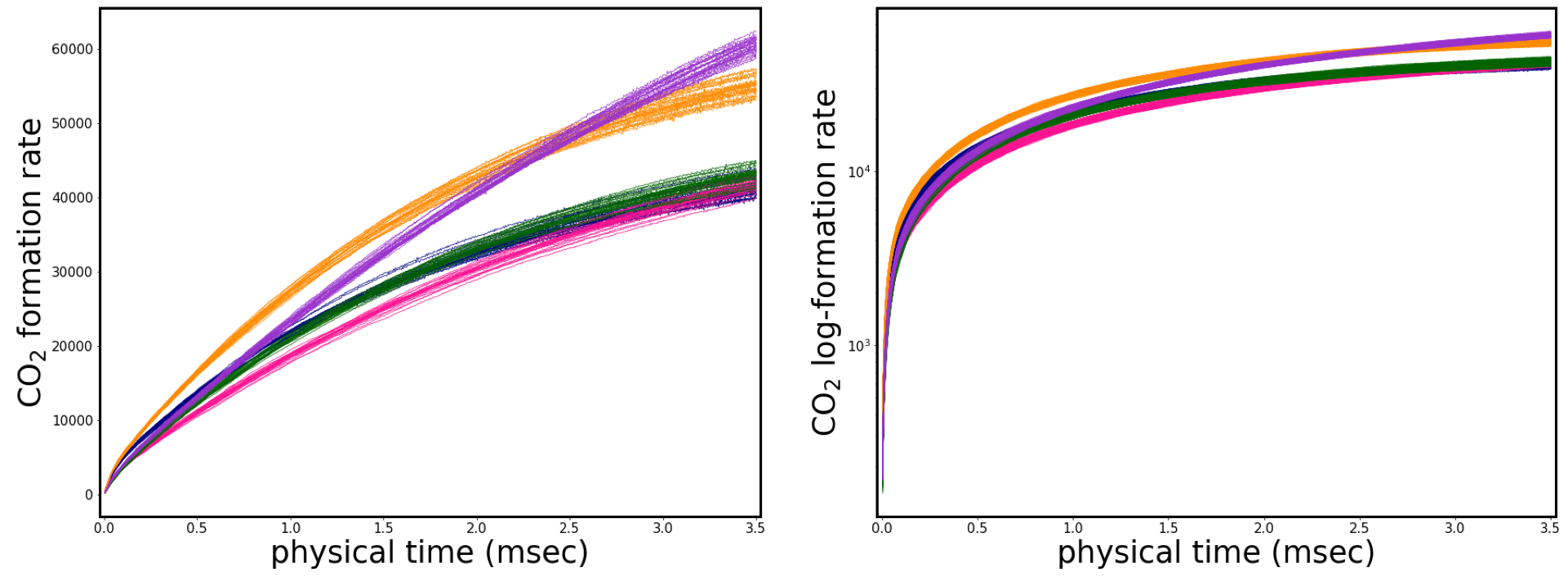}	
\caption{Production rate over time for CO$_2$. Each color corresponds to a different parameter configuration $\vlam^{(n)}$ where $n$ is fixed. Here the intrinsic stochasticity is highlighted by plotting individual trajectories which correspond to different random seeds $\omega$. The original trajectories are plotted on the left to illustrate the value separation within and between parametric samples. However, we curb the large value ranges by working with logarithms of the QoI, as shown on the right. }
\label{fig:ensembles}
\end{figure}

Figure~\ref{fig:ensembles} illustrates the time-evolution of the QoI for varying parameter values. Clearly, the production rate of the chemical system changes significantly with time. The effect of the parameters $\vlam$ is illustrated by the divergence of the sample solution medians to distinct trajectories over time, while the magnitude of intrinsic noise $\omega$ is indicated by the colored band surrounding each sample median. Looking ahead to global sensitivity analysis, the fractional variance contributions of intrinsic noise and parametric uncertainty are expected to be dominated by the latter over time, since the color bands seem well separated, except possibly where path crossing compresses the overall variance of the model. 

\subsection{Quantitative measures for results}
 
We illustrate construction of the joint PCE for $\eta_{l}(\vlam,\omega)$ and the KLPC surrogate for the production rate of CO$_2$ over time as our main QoI. The uncertain parameter vector is a set of 15 forward (log) reaction rate constants $\vlam \, = \, (\ln k_{1},...,\ln k_{15})$. Here we generate $N=1000$ parametric configurations sampled around nominal rates proposed in \cite{Dopking:2017} with an uncertainty scale factor of $r=1.50$ for each parameter. For each fixed parameter configuration we use our open-souce Kinetic Catalysis KMC software \verb#KinCat#~\cite{Daniels:2023} to generate $M=500$ replica solutions of the stochastic model on a \bt{lattice of $50 \times 50$ unit cells}. The initial condition of the surface is generated by selecting 25\% of the sites at random and scattering CO molecules and O atoms on these sites in even proportions.  

In this work, the joint PCE construction for $\eta_{l}(\vlam,\omega)$ is undertaken as a set of sequential steps. We first examine the quality of the stochastic and parametric PCEs separately to demonstrate that each step of the surrogate construction is sufficiently accurate. It follows that we must consider multiple measures of error related to the PCE goodness-of-fit, as the stochastic and parametric PCEs define fundamentally different approximations of the model outputs and cannot be evaluated in the same way. 

Whenever point estimates are appropriate, we compare surrogate predictions to the training data via the relative root mean-squared error (rRMSE) defined as
\begin{align} \label{eq:rRMSE}
E_{\text{\tiny rRMSE}} := \sqrt{ \frac{ \sum_{n=1}^{N} (v^{\text{\tiny train}}_{n} - v^{\text{\tiny pred}}_{n})^{2} }{ \sum_{n=1}^{N} \big(v^{\text{\tiny train}}_{n}\big)^{2} } }
\end{align}
where $v^{\text{\tiny train}}_{n}$ and $v^{\text{\tiny pred}}_{n}$ represent the training and predicted values, respectively.

\begin{figure}[t!]
\centering
\includegraphics[width=0.85\textwidth]{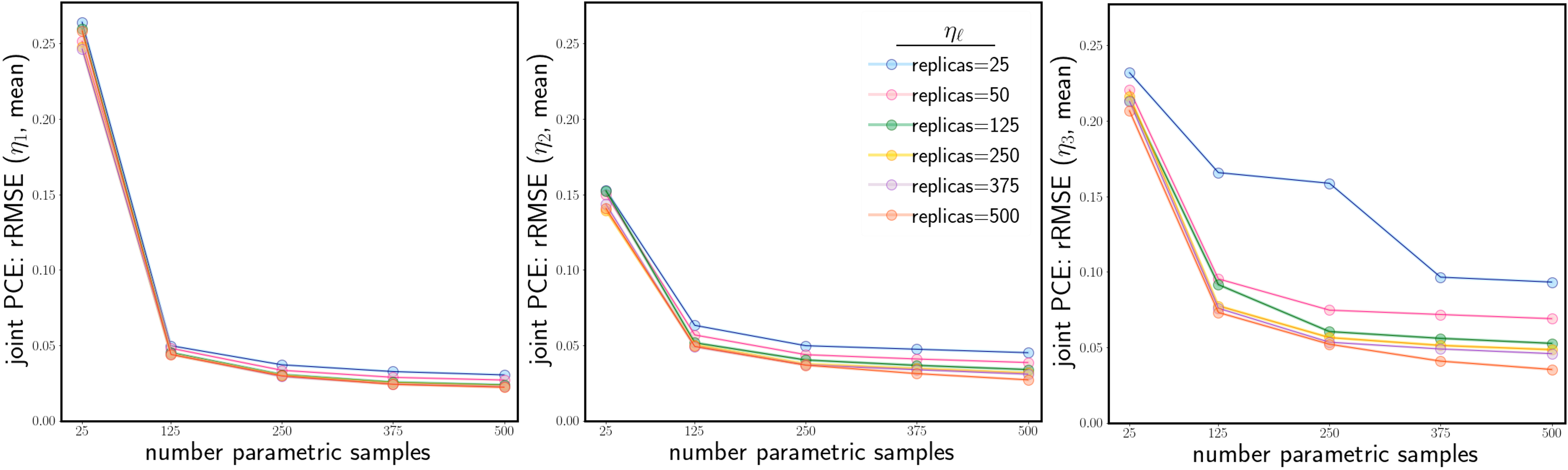}
\caption{Convergence plots with respect to training dataset size for the chemical kinetic example. Each line represents surrogates constructed using different numbers of sample replicas.}
\label{fig:convergence}
\end{figure}

\bt{We performed convergence studies on randomized 50/50 training/testing splits of the data, while the final surrogate was constructed from the combined set of training/testing samples. The convergence study also ensured we have enough training samples to be in a convergent regime. Figure~\ref{fig:convergence} demonstrates the reported test set rRMSE for each surrogate, plotted against the number of parametric training samples, while we vary both the number of parametric samples \emph{and} the number of sample replicas used in the joint PCE construction. The results in this figure illustrate that most of the reduction of the rRMSE in the dominant KLE terms $\eta_1$ and $\eta_2$ is achieved with roughly $N_p\approx 125$ parametric samples and $N_s \approx 50$ replicas per parameter. The error in the noisiest element $\eta_3$ required a somewhat larger number of samples to roughly approach its plateau.}

\subsection{KLE for production rate over time}

The first step is to find a suitable compact representation of the time-evolving system. Random fields are well represented by the KLE, and highly correlated fields can usually withstand substantial truncation of the expansion without suffering significant loss of information. As noted in Section~\ref{sec:field}, the rate of spectral decay dictates how early truncation can occur. That is, the rate of variance accumulation in the KLE terms is tied to the rate of decrease in the eigenvalue magnitudes, since the captured variance is given by the sum of the eigenvalues. 

\begin{figure}[h!] 
\centering
\includegraphics[width=0.75\textwidth]{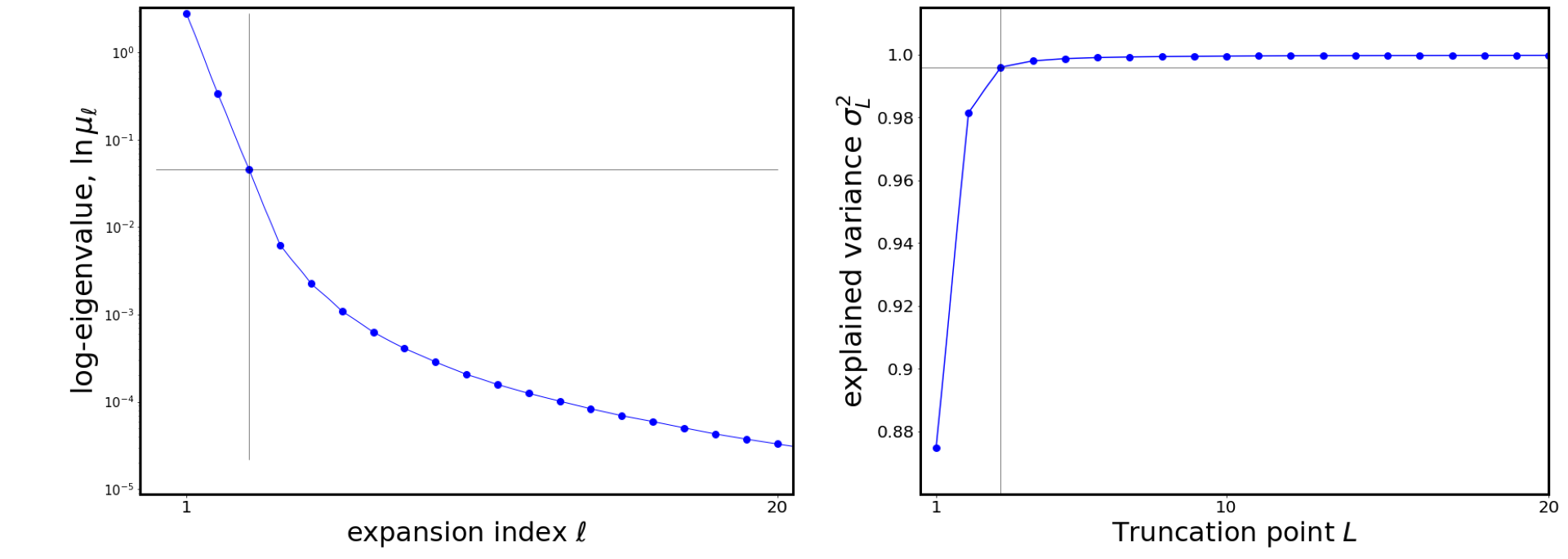}	
\caption{Spectral decay is shown by plotting the (log) eigenvalue against expansion index $l$ (left). Accumulation of explained variance in the KLE is demonstrated by plotting the explained variance given by Eq.~\eqref{eq:fracvar} against proposed truncation points $L$ (right). The crossing point of the gray lines indicates where the KLE captures at least 99\% of the explained model variance.}
\label{fig:kle}
\end{figure}

Figure \ref{fig:kle} clarifies the relationship between the spectral decay and accumulation of explained variance. Here we take advantage of the ordering of the eigenvalues and plot the (log) eigenvalues against the expansion index $l$. The early eigenvalues exhibit orders-of-magnitude value changes that result in gaps in the spectrum and large changes in the accumulated variance. As the rate of spectral decay slows, the accumulation of variance similarly levels off. Typically the truncation point is chosen to preserve some proportion of the output variance.

\begin{figure}[ht!]
\centering
\includegraphics[width=\textwidth]{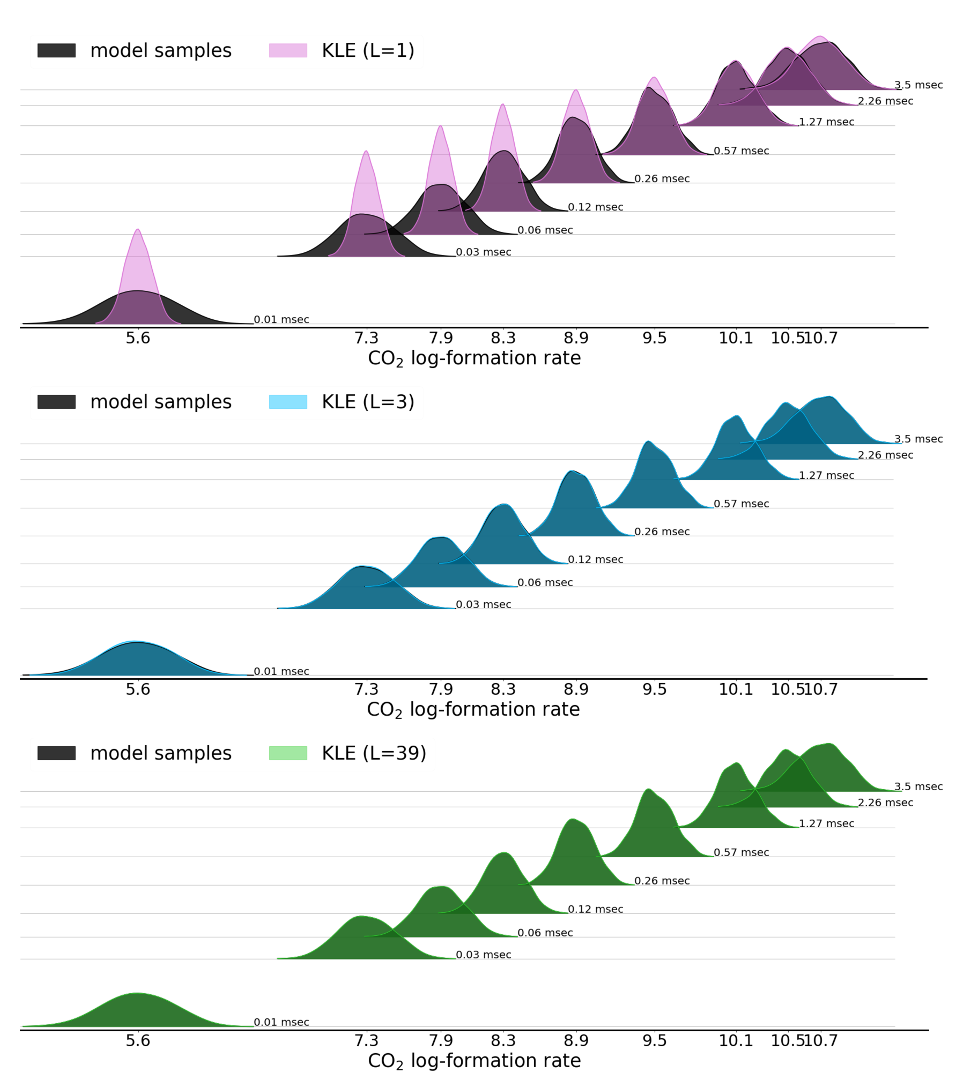}
\caption{PDFs of the model samples (gray) and KLE representation over time at different truncation points: $L=1$ (pink), $L=3$ (blue), and $L=39$ (green). Note that the vertical distances between PDFs is a log-scale of the simulation time to improve visibility of PDFs. The values along the $x$-axis indicate the mean of each PDF over time. }
\label{fig:kle_modes_2}
\end{figure}

\begin{figure}[t!]
\centering
\includegraphics[width=0.8\textwidth]{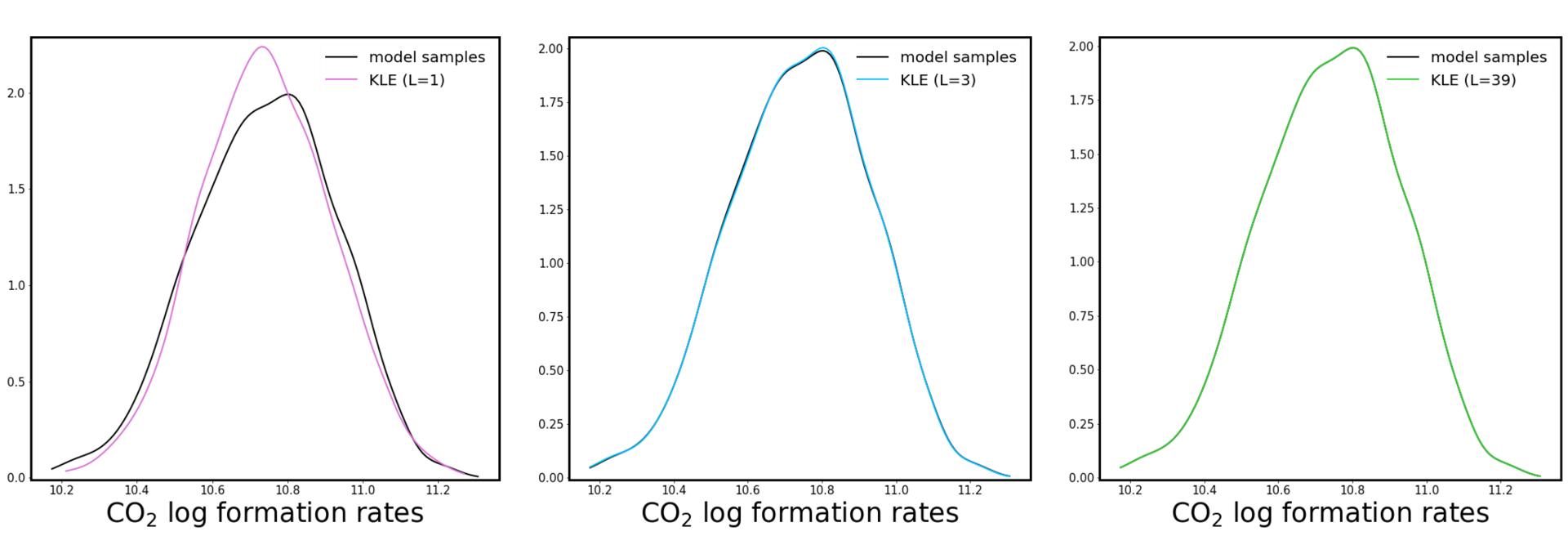}
\caption{PDFs of the model samples at the final simulation time (gray) and KLE representation at different truncation points. The number of terms in the KLE increases from left to right.}
\label{fig:kle_modes}
\end{figure}

The effect of the truncation point $L$ on the error in the KLE of the training samples is illustrated in Figures~\ref{fig:kle_modes} and \ref{fig:kle_modes_2}. PDFs of the \emph{combined} stochastic-parametric training samples are plotted in gray while KLE reconstructions of these data are plotted in pink ($L=1$), blue ($L=3$), and green ($L=39$). Note that $L=1$ produces the most compact representation of the data possible, while $L=3$ and $L=39$, according to Figure~\ref{fig:kle}, capture at least 99\% and 99.99\% of the total explained variance, respectively. 

Figure \ref{fig:kle_modes_2} illustrates the quality of the KL representation of these models relative to the training data over time, while Figure \ref{fig:kle_modes} compares the PDFs of the QoI and its KL representation at the final KMC simulation time once the trajectories have (approximately) reached equilibrium. Increasing the number of KLE terms results in a more accurate KL approximation across all simulation times, but comes at the expense of higher-dimensional representations of the data. The key is to find a reasonable balance between the accuracy of the model and the number of features we must construct surrogates for. From this comparison, we select the KLE with $L=3$ terms. This captures slightly more than 99\% of the explained variance and produces a reasonably accurate approximation of the training data, while curbing the number of elements that must be estimated with joint PCEs. 

\begin{figure}[ht!] 
\centering
\includegraphics[width=0.75\textwidth]{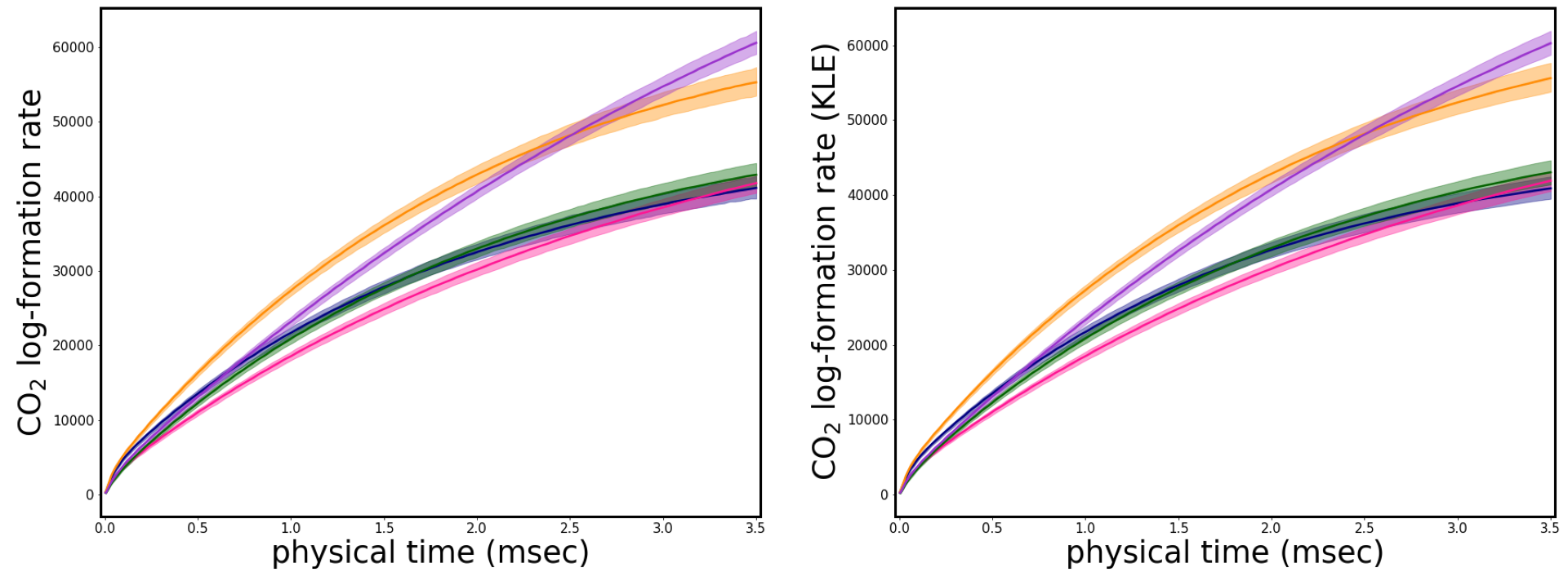}
\caption{Production rate over time for CO$_2$. Each color corresponds to a different parameter configuration $\vlam^{(n)}$, $n$ fixed. The ensemble medians are plotted as thick lines and the $5$th/$95$th quantiles as a band around the median. The original model samples (left) are compared to their KLE representation when $L=3$ (right). We see that the KLE solution trajectories reproduce the correct behavior over time for this level of truncation. }
\label{fig:kle_ensembles}
\end{figure}

We compare the selected KLE representation to the training samples in Figure~\ref{fig:kle_ensembles} and see that the KLE solution trajectories reproduce the correct behavior over time. The medians of the solution trajectories are plotted as thick, colored lines for individual parametric samples $\vlam^{(n)}$, $n$ fixed, and the variation in samples indicated by the colored band surrounding each median which represent the $5$th/$95$th percentiles. 

\subsection{Stochastic PCE}

With a reasonable low-dimensional representation of the random field data in hand, we now seek stochastic PCE representations of the random coefficients $\eta_{l}(\vlam,\omega)$ for each $l=1,2,3$. The PCEs take the form given in Eq. \eqref{eq:stochastic_pce}.

We limited the stochastic PCEs to first-order expansions of the germ \bt{(intrinsic noise)} data, as higher-order expansions \bt{for the noise} appeared to offer little gain in accuracy for the QoI while rapidly increasing the number of parametric surrogates needed to construct the joint PCE. Since the sample replicas capture intrinsic model noise and are (approximately) normally distributed, we use the Gauss-Hermite polynomials as the expansion basis.

Due to lack of control over $\omega$ and $\bm{\zeta}$, the training samples and the stochastic PCE predictions of $\bm{y}(t)$ are only guaranteed to agree in distribution. Therefore, we assess the overall performance of the stochastic PCEs by comparing statistics of the training model simulations and the PCE-predicted random coefficients. For each fixed $\vlam^{(n)}$, we take sample replicas to find the sample means and standard deviations for PCE predictions of $\bm{\eta}^{\text{\tiny PCE}}(\vlam^{(n)},\omega)$. We then plot each component of the predicted means (standard deviations) against the true sample means (standard deviations) in parity plots. 

\begin{figure}[t!] 
\centering
\includegraphics[width=0.85\textwidth]{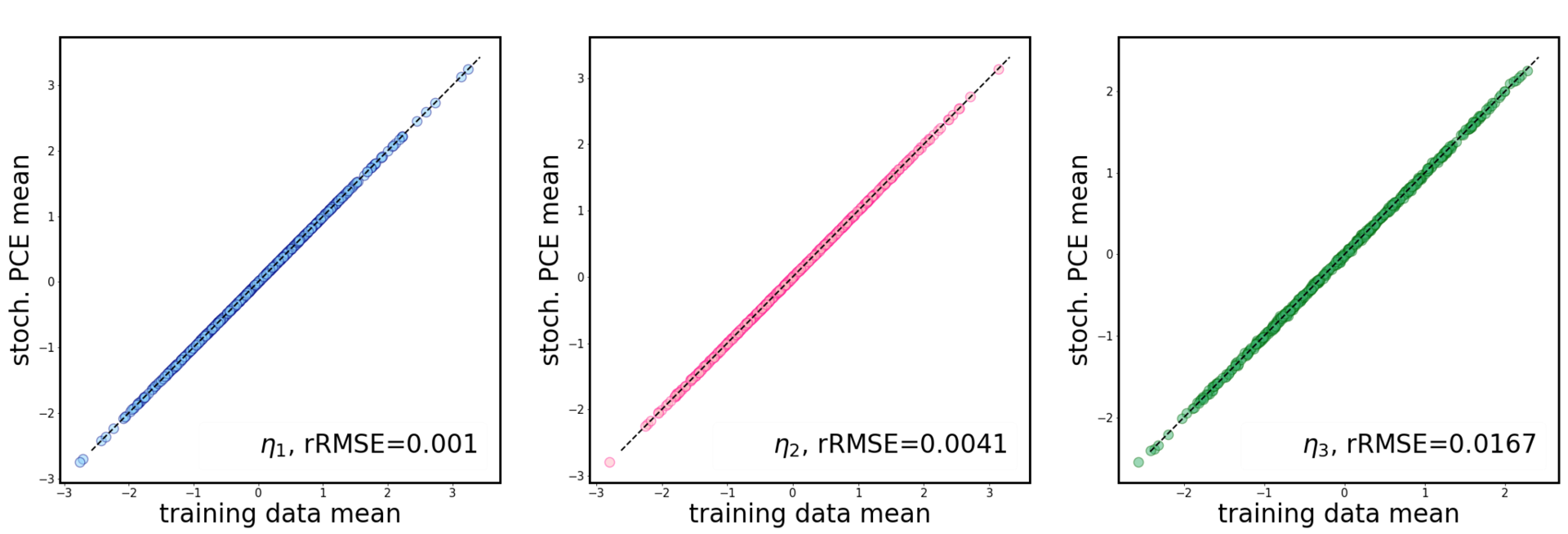}
\includegraphics[width=0.85\textwidth]{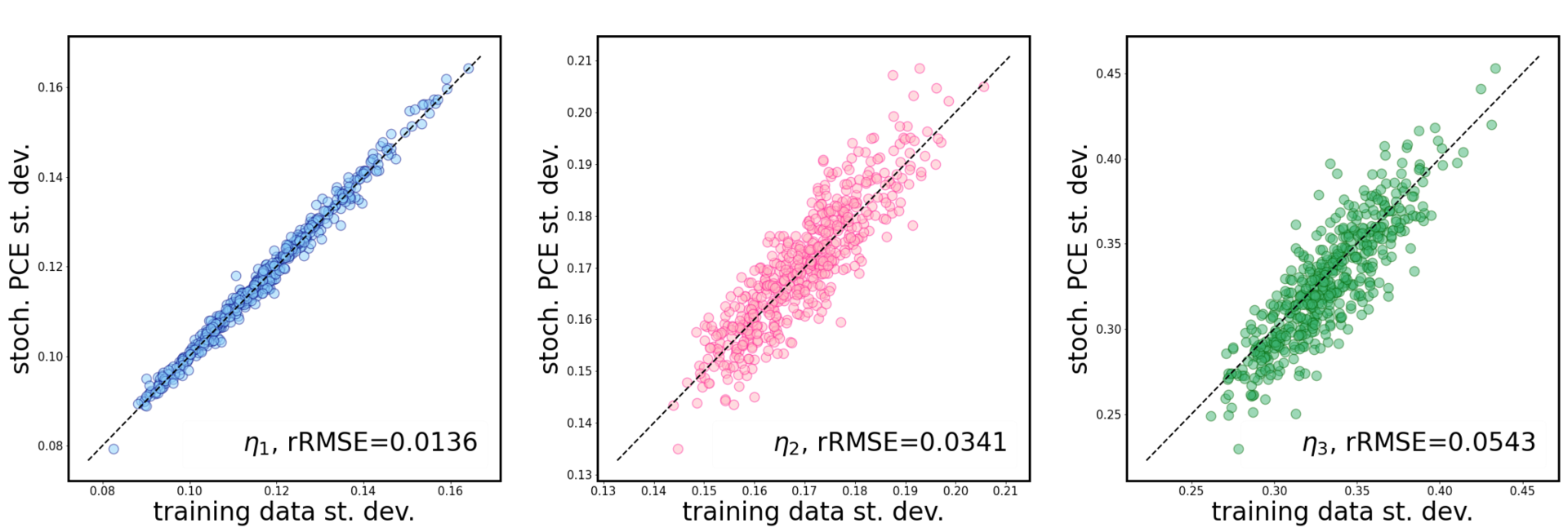}
\caption{Parity plots of the mean (top row) and standard deviation (bottom row) of the random KLE coefficients $\eta_{l}(\vlam,\omega)$ predicted by the stochastic PCE. The rRMSE for the mean and standard deviation of the set of combined samples are $0.0043$ and $0.0272$, respectively. }
\label{fig:dm_stoch_pce}
\end{figure}

Figure \ref{fig:dm_stoch_pce} demonstrates this comparison, indicating that the stochastic PCEs predict the sample means with high accuracy for all three KLE coefficient variables $\eta_1(\vlam,\omega), \eta_2(\vlam,\omega)$, and $\eta_3(\vlam,\omega)$. A similar comparison of the predicted standard deviations reveals that the stochastic PCEs predict the variance of the distributions well, on average, but the spread of the predictions increases with the amount of noise in the random coefficient (see Figure \ref{fig:dm_stoch_pce}). We reduce the spread by increasing the number of samples in the training dataset. Further, the bandwidth used to construct the Rosenblatt maps must be chosen with care to avoid severely biasing the predicted standard deviations of samples. Particularly, a bandwidth based on Silverman's rule-of-thumb~\cite{Silverman:1986} is not optimal for moment approximation and leads to systematic over-estimation of predicted variances in the stochastic samples. Here the bandwidth was chosen empirically, and automated optimization of these values will be addressed in future work. 

\subsection{Parametric PCE}

Construction of stochastic PCEs for the 
$\eta_{l}(\vlam,\omega)$,  
\begin{align} \label{eq:stoch_pce_results}
\eta_{l}(\vlam,\omega) 
	= \sum_{s=0}^{3} \bm{z}_{ls}(\vlam) \Psi_{s}(\bm{\zeta}), \quad l=1,2,3
\end{align}
yields model outputs $\bm{z}_{ls}(\vlam)$ that directly correspond to the parametric input samples $\vlam$. Here the subscript $l$ associates explicitly a particular $\eta_l$ with the coefficients of its representation, while subscript $s$ indicates the particular element of the sum. We now construct parametric PCEs to represent this input-output mapping.

Since the elements of $\bm{\lambda}$ are assumed to be distributed uniformly on $[\ln \bm{k}_{\text{\tiny nom}} - \ln \bm{r}, \ln \bm{k}_{\text{\tiny nom}} + \ln \bm{r}]$ \bt{and we seek uniform point accuracy of the surrogate on the domain}, we use Legendre polynomials as the expansion basis and obtain the simple rescaling relationship as a first-order Legendre-Uniform PCE,
\begin{align} 
\lambda_{i} 
	= \ln k_{\text{\tiny nom,} \,  i} + \xi_{i} \ln r_{i} , \text{ for } i=1,\dots, \tilde{d}.
\end{align}
As described in Section~\ref{sec:methods}, we then construct a parametric PCE of the form Eq.~\eqref{eq:coefficient_pce},
\begin{align}
\bm{z}_{ls} 
	= \sum_{p=0}^{P_{ls}-1} \bm{a}_{lsp} \Psi_{p}(\vxi).
\end{align}
This gives the stochastic PCE coefficients $\bm{z}_{ls}$ as functions of the rescaled input, or germ, $\bm{\xi}$.

The parametric PCEs are chosen to have at most second-order expansions of the coefficient data $\bm{z}_{ls}$, and the optimal orders for each $s$ are selected via model evidence which is analytically available for this regression step~\cite{Sargsyan:2016}. We reduce the number of expansion terms in the parametric PCE by estimating the coefficients with Bayesian Compressive Sensing (BCS), which adaptively finds a sparse, optimal basis set for the PCE ~\cite{Babacan:2010, Luthen:2021, Sargsyan:2014}. We examined the accuracy of the parametric PCEs using randomized \bt{50/50} training/testing splits of the data with the rRMSE metric on coefficient approximation, followed by a final training of the parametric surrogate using all samples. 

Unlike the stochastic PCEs, the parametric PCE preserves pointwise mappings between the PCE germ $\bm{\xi}$, parametric inputs $\bm{\lambda}$, and $\bm{z}_{ls}(\bm{\lambda})$ coefficient values, hence we do not need to resort to moment comparisons and can examine pointwise parity plots of the true and PCE-predicted outputs. 
 \begin{figure}[t!] 
\centering
\includegraphics[width=0.85\textwidth]{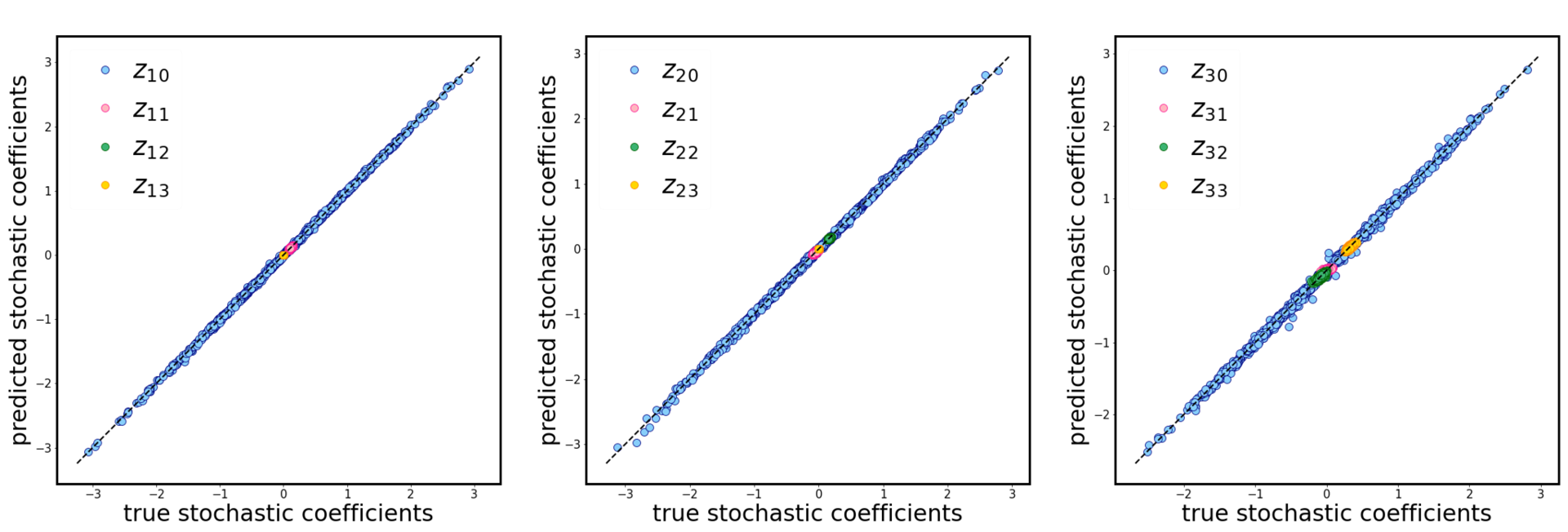}	
\caption{Parity plots of parametric PCE predicted coefficients $\bm{z}_{ls}^{\text{\tiny PCE}}(\bm{\lambda})$ constructed in Eq.~\eqref{eq:coefficient_pce} vs true coefficients $\bm{z}_{ls}(\bm{\lambda})$ found in Eq.~\eqref{eq:stochastic_pce2} for all ensemble members. The coefficients $\bm{z}_{l}(\bm{\lambda})$ are grouped by random KLE element $\eta_{l}(\vlam,\omega)$.}
\label{fig:dm_param_pce}
\end{figure}
Figure~\ref{fig:dm_param_pce} illustrates the approximation quality of each parametric PCE coefficient by plotting the predicted values $\bm{z}_{ls}^{\text{\tiny PCE}}(\vlam)$ against the true $\bm{z}_{ls}(\vlam)$ for each $s=0,1,2,3$. 
We reduce the total number of plots needed to show this result by collecting all $\bm{z}_{ls}$ coefficients of a particular $\eta_{l}$ together in one plot.

The close fit of the models to the training data for each coefficient is supported by the computed rRMSE for each collection of coefficients: rRMSE$(\bm{z}_{1})=0.019$,  rRMSE$(\bm{z}_{2})=0.028$, rRMSE$(\bm{z}_{3})=0.048$.

\subsection{Joint PCE, KLPC, and Global Sensitivity Analysis}

Finally, we assemble the joint PCE predictions sequentially. We substitute the parametric PCE results for $\bm{z}_{ls}$ into the stochastic PCE to obtain the joint PCE predictions of $\eta_{l}(\vlam,\omega)$:
\begin{align}
\eta_{l}(\vlam,\omega)
	= \sum_{s=0}^{3}\bigg( \underbrace{ \sum_{p=0}^{P_{ls}-1} \bm{a}_{lsp} \Psi_{p}(\vxi)  }_{\bm{z}_{ls}} \bigg) \Psi_{s}(\bm{\zeta})
\end{align}
These are further folded into Eq.~\eqref{eq:klpc}, the KLE, to obtain physical QoI samples of the form of Eq.~\eqref{eq:klpc_2} from the surrogate. 

\begin{figure}[t!]
\centering
\includegraphics[width=0.4\textwidth]{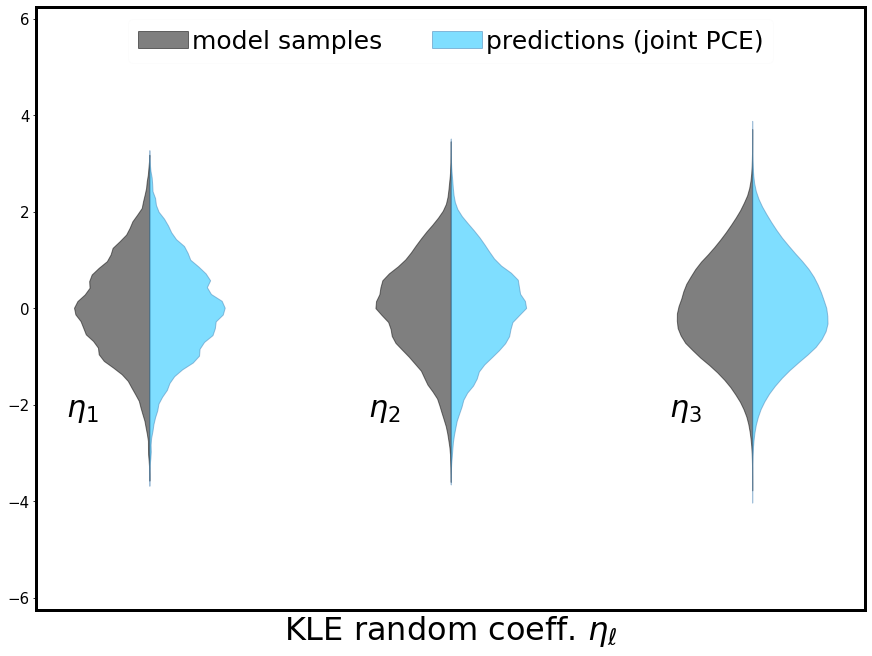}
\caption{Violin plots demonstrating the accuracy of the joint PCE predictions in spectral space.}
\label{fig:violin_jnt}
\end{figure}

\begin{figure}[h!]
\centering
\includegraphics[width=0.75\textwidth]{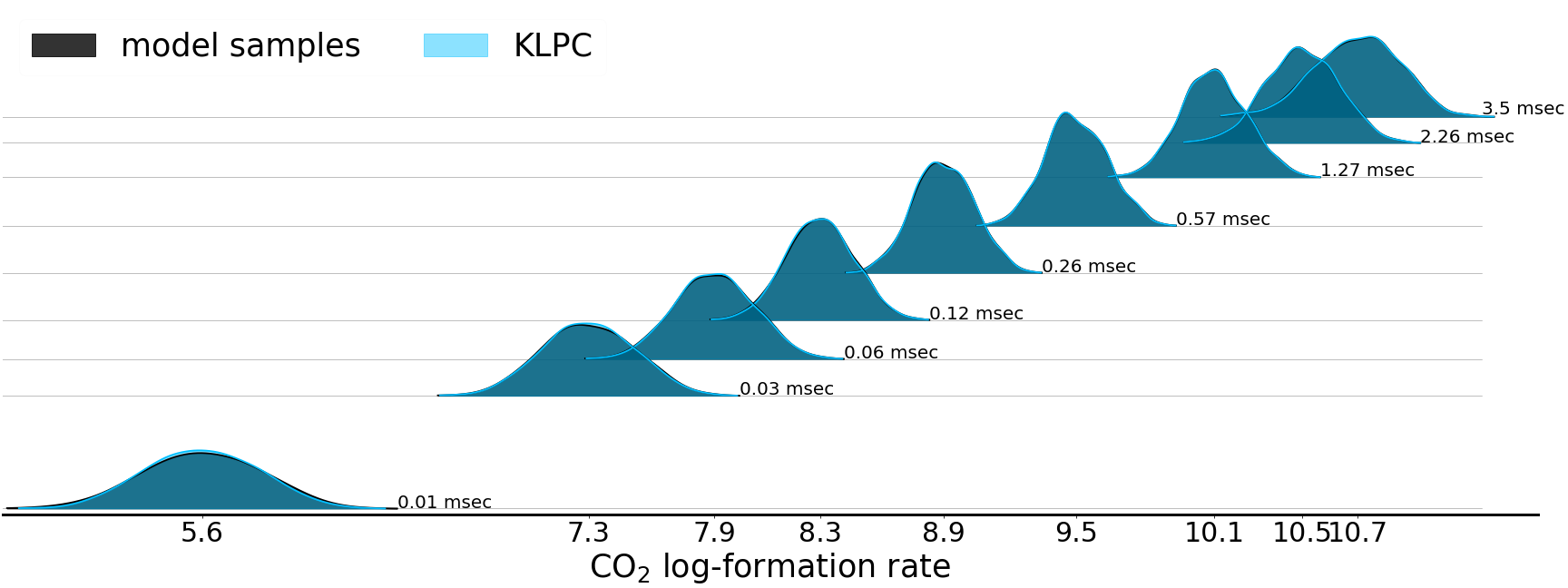}
\caption{Ridge plots compare the combined KLPC predictions over time to the training samples. The values along the $x$-axis indicate the mean of each PDF over time.} 
\label{fig:ridge_klpc}
\end{figure}

As described in Section~\ref{sec:field}, the KLPC surrogate is itself a PCE and its quality is dependent upon both the quality of the samples of $\eta_{l}(\vlam,\omega)$ generated by the joint PCE and the number of KLE modes included in the initial low-dimensional representation of the random fields. In Figures \ref{fig:violin_jnt}-\ref{fig:ridge_klpc} we compare these surrogate predictions to the training data on the set of combined stochastic-parametric samples. \newpage

Examining the quality of both the joint PCE predictions of $\eta_{l}(\vlam,\omega)$ (see Figure~\ref{fig:violin_jnt}) and the KLPC predictions of $f(\vlam,\omega; \vt)$ (see Figure~\ref{fig:ridge_klpc}) show that the surrogate is accurate on both the spectral space of KLE elements and physical space of QoIs. 

As a bi-product of the KLPC construction~\eqref{eq:klpc_2}, one can extract global sensitivity indices and attribute the variance in the output QoI to either the uncertainties in the parameters or intrinsic noise from the 
\begin{figure}[b!] 
\centering	
\includegraphics[width=0.425\textwidth]{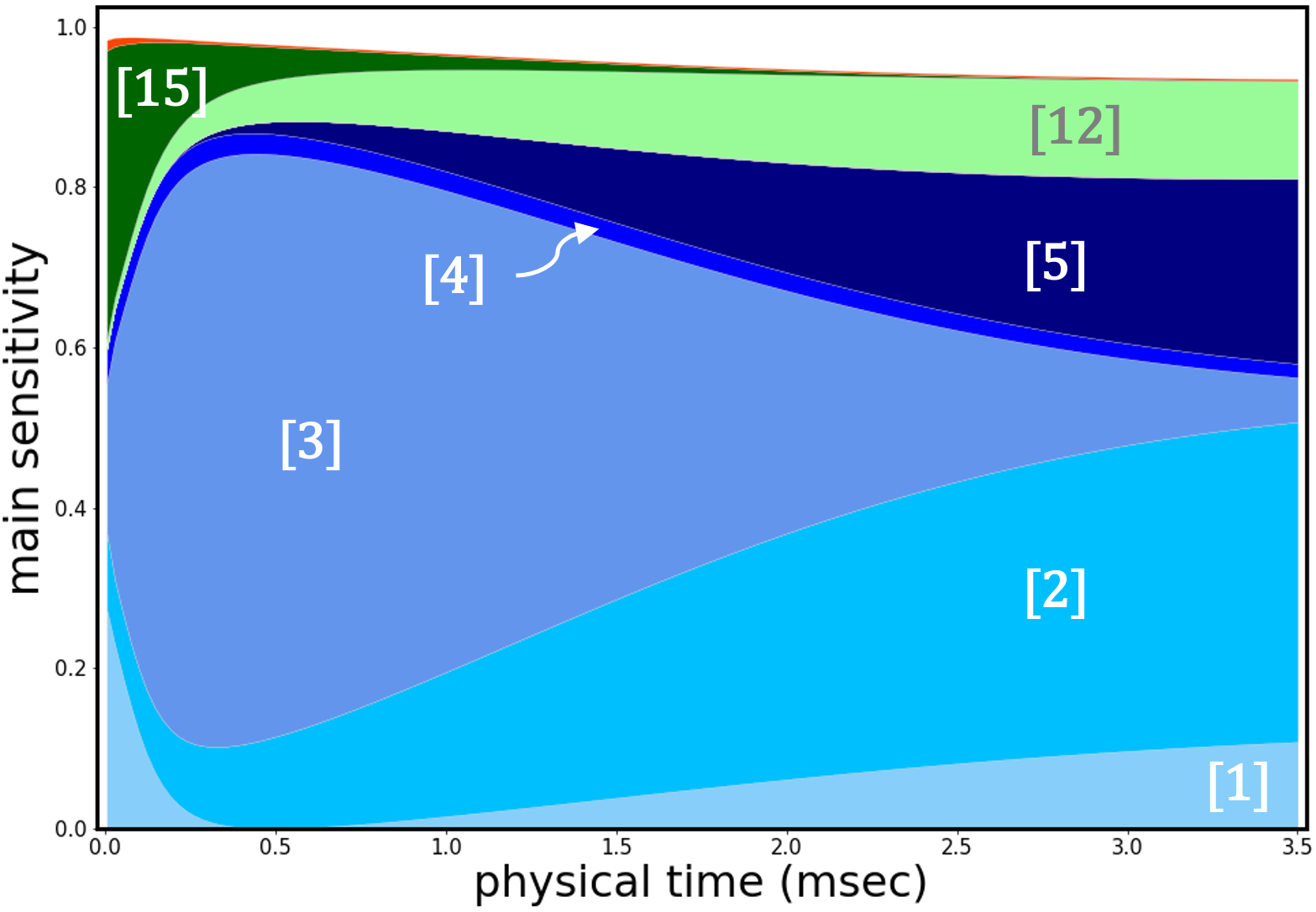}	
\caption{Sobol indices over time for uncertain rate constants $k_{i}$ of processes described in Table~\ref{tab:rxns}. Only adsorption rates (blues), certain formation rates (greens), and noise (red) contribute appreciably to the sensitivity analysis. Sensitivity indices of these most important parameters are labeled by associated process.}
\label{fig:gsa}
\end{figure}
stochastic model. As illustrated in Figure~\ref{fig:gsa}, we find the adsorption rates to be the largest contributors to the output variance for all times in the simulation, although the importance of the formation rates increases as the lattice fills and the number of formation events stabilizes. The Sobol index from the intrinsic model noise is a group sensitivity that is found by summing the effects of all elements of $\bm{\zeta}$.

The parametric uncertainties overall seem to dominate the intrinsic noise as one would expect intuitively based on Figure~\ref{fig:ensembles}. Effectively, the high parametric uncertainty washes out the influence of intrinsic noise in the system. Meanwhile, the highest noise contribution to the output uncertainty occurs where the number of CO$_2$ formation events rapidly increases. Here, path crossing compresses the overall variance of the model output due to the parameters and increases the sensitivity of the QoI to the influence of intrinsic noise.

Finally, for this GSA result, we point out the effect of cross-interactions introduced by the second order expansion terms. It is known that main sensitivities will sum to one if cross terms are absent from the polynomial expansions.  When cross terms are present, they siphon off a fraction of the main sensitivities and the sum of the main indices is less than one. The results in Figure~\ref{fig:gsa} indicate that the contribution of cross terms is small initially, growing slightly larger in time.

\begin{figure}[t!] 
\centering
\includegraphics[width=0.9\textwidth]{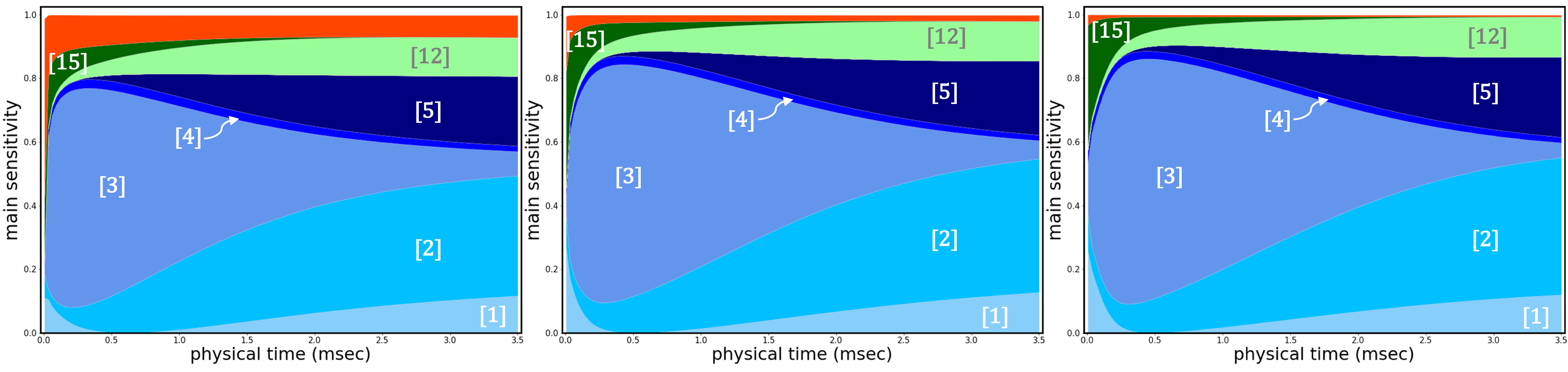}	
\caption{Sobol indices over time for uncertain adsorption rates (blues), formation rates (greens), and noise (red) of processes described in Table~\ref{tab:rxns}. The lattice sizes increases from left-to-right and have dimensions $n_{c}=25$ (left), $n_{c}=50$ (center), and $n_{c}$=100 (right). Sensitivity indices of the most important parameters are labeled by associated process.}
\label{fig:gsa_comparison}
\end{figure}

One strength of this approach for stochastic surrogate construction is the following: the effect of intrinsic noise on the model output can be systematically quantified and thereby controlled. That is, we can use the joint PCE (KLPC) construction as a model selection tool that guides, for example, the ensemble size chosen to reduce intrinsic noise when estimating integrated QoIs or the size of the lattice used in a KMC simulation to generate data samples. This is demonstrated in Figure~\ref{fig:gsa_comparison}, where the Sobol indices over time are plotted for KLPC surrogates constructed from data generated from KMC simulations which differ only in the lattice size $n_{c}$. As the lattice size increases, the presence of intrinsic noise (plotted with red) is diminished until it is negligible. From this we conclude that the largest lattice, with dimension $n_{c}=100$, is optimal in that it is the smallest lattice which allows nearly noise-free samples to be generated for the catalytic system under investigation.

In Figure \ref{fig:gsa_comparison}, we ensure consistency amongst the training datasets generated with KMC by using the same set of parametric inputs $\vlam$ when constructing trajectories. We further initiate each simulation from the same set of initial random seed values. Thus, the only significant difference in the results comes through the influence of the lattice size, $n_{c}$. To reduce the computational effort of generating training data for this comparison, we reduced the parametric uncertainty and used a smaller number of parametric inputs samples than in the previous results. 

\subsection{Using the KLPC to obtain sensitivity indices for log-normal rates}

Constructing the surrogate model from independent, log-uniform rate samples ensures that the joint PCE surrogate is uniformly accurate over the hypercube $[a_{i}, b_{i}], i=1,\dots,15$. Here the bounds are given from the nominal rates and uncertainty scale factors via $\ln k_{\text{\tiny nom},i} \pm \ln r_{i}$. However, chemical reaction rates are best described with log-normal distributions. To obtain more physically meaningful GSA results, we make use of the following procedure, presuming independent log-normal rate uncertainties. 
\bi
\item[1.] With the surrogate model Eq.~\eqref{eq:klpc}, we generate $N^{\prime}$ samples from the standard normal germ $\bm{\nu} \sim \mathcal{N}(0,1)$. These can be mapped to physical rate values via $\ln k_{i} = \mu_{i} + \nu_{i} \sigma_{i}$ -- so that $\ln k_{i} \sim \mathcal{N}(\mu_{i}, \sigma_{i}^{2})$ -- where $\mu_{i}$ and $\sigma_{i}$ are the mean and standard deviation of the $i$th process rate, respectively. Denote the collection of log-normally distributed rates $\vtheta = [\ln k_{1}, \dots, \ln k_{15}]$.
	\bi
	\item[(a)]
	The mean is naturally given by the nominal rate values $\mu_{i} = \ln k_{\text{\tiny nom}, i}$.

	\item[(b)] The sample standard deviation is obtained from confidence intervals. That is, given a sample set of size $N^{\prime}$ with mean $\ln k_{\text{\tiny nom},i}$ and standard deviation $\sigma_{i}$, there is some probability $p(z)$ that a value $\ln k_{i}$ lies within $z$ standard deviations of the mean:
	\begin{align*}
	\ln k_{i} = \ln k_{\text{\tiny nom},i} \pm \frac{z \sigma_{i} }{\sqrt{N^{\prime}}}.
	\end{align*}
	If we wish to ensure that $\ln k_{i} \in [\ln k_{\text{\tiny nom},i} - \ln r_{i} ,\ln k_{\text{\tiny nom},i} + \ln r_{i} ]$ with probability $p(z)$, then set $\ln r_{i} = \frac{z \sigma_{i} }{\sqrt{N^{\prime}}}$. Solving for $\sigma_{i}$ gives the sample standard deviation we should choose:
	\begin{align} \label{eq:std}
	\sigma_{i} = \sqrt{N^{\prime}} \, \frac{\ln r_{i}}{z}.
	\end{align}
	Note that any sample $\ln k_{i}$ may fall outside of the desired interval with probability $1-p(z)$. Standard normal tables can provide the one-to-one correspondence between $p$ and $z$, and one can chose $z$ according to a given threshold value of $p$. With $p$ being close to one, we still obtain a reasonable estimate for the Sobol indices for GSA with only small number of normal samples falling outside the uniform range. While the GSA is accurate enough, the model predictions of the QoI at these inputs is untrustworthy as they are outside the range of validity of the pre-constructed surrogate.
	\ei 
	
\item[2.] With the surrogate model Eq.~\eqref{eq:klpc}, germ samples $\bm{\nu}$, and associated physical rate values $\vtheta$ in hand, we evaluate the surrogate model at values of $\vtheta$ to obtain associated samples of the QoI,
\begin{align} \label{eq:gtheta}
g(\vtheta, \omega; \vt) 
	= f^{\text{\tiny PCE}}(\vxi(\vtheta), \vzeta; \vt).
\end{align}
As before, we transform independent elements of the input rates $\theta_{i}$ to the compact interval of support $[-1,1]$ via rescaling $\theta_{i} = \ln k_{\text{\tiny nom},i} + \xi_{i} \ln r_{i}$. These transformed values $\vxi (\vtheta)$ are evaluated in the surrogate, Eq.~\eqref{eq:gtheta}.

\item[3.] Now, given the new dataset of log-normal rates and associated QoIs, $\{ \vtheta, g(\vtheta, \omega; \vt)\}$, we can construct a \emph{second} joint PCE surrogate such that \emph{both} the stochastic and parametric PCEs utilize a \emph{standard normal} germ, i.e.
\begin{align} \label{eq:klpc_nu}
g(\vtheta, \omega; \vt) 
	\approx g^{\text{\tiny PCE}}(\bm{\nu}, \vzeta; \vt)
	= \sum_{j=0}^{J-1} \bigg( \underbrace{ f_{0}(\vt) \delta_{0j} + \sum_{l=1}^{L} \tilde{\bm{b}}_{lj} \sqrt{\mu_{l}} \phi_{i}(\vt) }_{\bm{c}^{\text{\tiny KLPC}}(\vt)}  \bigg) \Psi_{j}(\bm{\nu}, \vzeta).
\end{align}
The new surrogate Eq.~\eqref{eq:klpc_nu} is used to propagate uncertainty from the physical rates $\vtheta$ and Sobol indices are obtained from the PCE coefficients $\bm{c}^{\text{\tiny KLPC}}(\vt)$:
\begin{align}  \label{eq:gsa}
S_{i} &= \frac{1}{\mathbb{V}[g(\vt)]} \sum_{j \in \mathcal{J}_{S_{i}}} c_{j}(t)^{2} \vert \vert \Psi_{j} \vert \vert^{2}
         = \frac{1}{\mathbb{V}[g(\vt)]} \sum_{j \in \mathcal{J}_{S_{i}}} \small{\bigg(\sum_{l \leq L} \tilde{\bm{b}}_{l j} \sqrt{\mu_{l}} \, \phi_{l}(\vt)  \bigg)^{2}}  \vert \vert \Psi_{j} \vert \vert ^{2}.
\end{align}
\ei
As before, $\vert \vert \Psi_{j} \vert \vert^{2}$ is defined analogously to Eq.~\eqref{eq:normed_psi}.

In summary, given nominal rates and uncertainty scale factors, we can generate log-normal reaction rate samples $\vtheta$, and then evaluate these rates in the original Legendre-Uniform KLPC surrogate to quickly obtain samples of the QoI. As noted previously, the generative nature of the KLPC allows us to generate an arbitrary number of sample replicas for each fixed $\vtheta^{(n)}$.

\begin{figure}[ht!]
\centering
\includegraphics[width=0.9\textwidth]{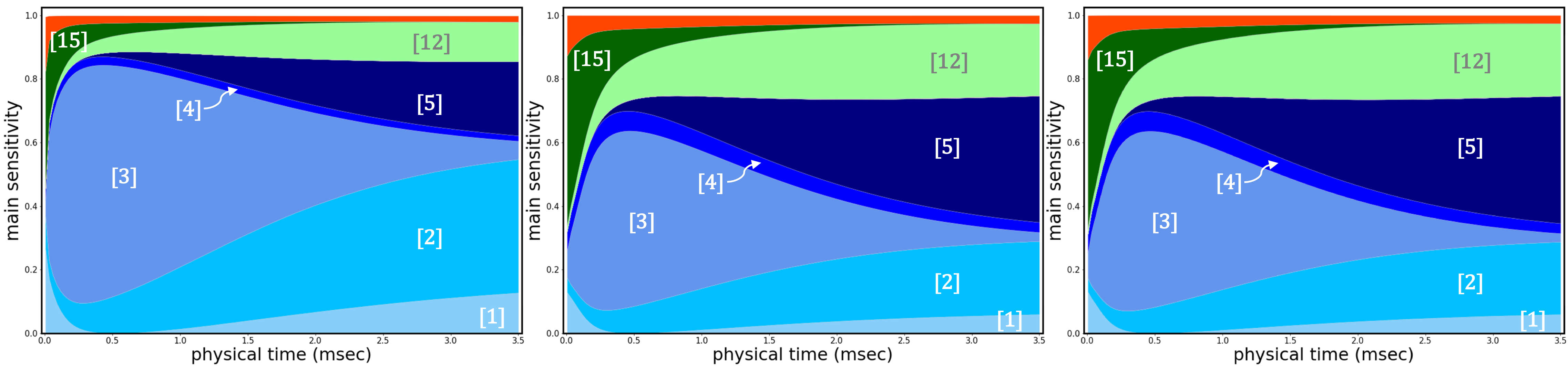}
\caption{Comparison of Sobol indices over time for KLPC surrogates constructed from three different training datasets. Left:~Training data are generated with KMC and the input parameters $\vlam$ follow log-uniform distributions. Right:~Training data are generated with KMC and the input parameters $\bm{\theta}$ follow log-normal distributions. Center:~Training data are generated for the same log-normal input parameters $\bm{\theta}$ by sampling the uniformly-accurate KLPC. Sensitivity indices of the most important parameters are labeled by associated process as given in Table~\ref{tab:rxns}.}
\label{fig:gsa_compare}
\end{figure}

We use these surrogate-generated samples of the input/output data $\{\vtheta,f^{\text{\tiny PCE}}(\vxi(\vtheta),\omega;\vt)\}$ to construct a Gauss-Hermite KLPC Eq.~\eqref{eq:klpc_nu} which exclusively uses standard normal germs throughout the construction. The coefficients $\tilde{\bm{b}}$ of this normal-based surrogate are then used to compute Sobol indices over time. We plot these indices in Figure \ref{fig:gsa_compare} and compare to the Sobol indices obtained from the original Legendre-Uniform surrogate and to the Sobol indices obtained from directly constructing a Gauss-Hermite PCE-based surrogate on KMC samples generated with the same log-normal rates. \bt{Regardless of the distribution of the input parameters, the identity of and significance of the sensitive parameters are preserved over time.}
\bt{We reiterate this} GSA result \bt{(center of Figure \ref{fig:gsa_compare}) is} obtained from \bt{a \emph{Gauss-Hermite}} KLPC trained on data sampled from the original \bt{\emph{Legendre-Uniform}} surrogate, \bt{yet it} does not produce an appreciable amplification of noise. \bt{This example highlights the utility of choosing the most general surrogate construction possible for the given data, as it enabled multiple explorations of the chemical kinetic system without requiring additional KMC samples.} 
\subsection{A bimodal example}

\begin{figure}[h!]
\centering
\includegraphics[width=0.55\textwidth]{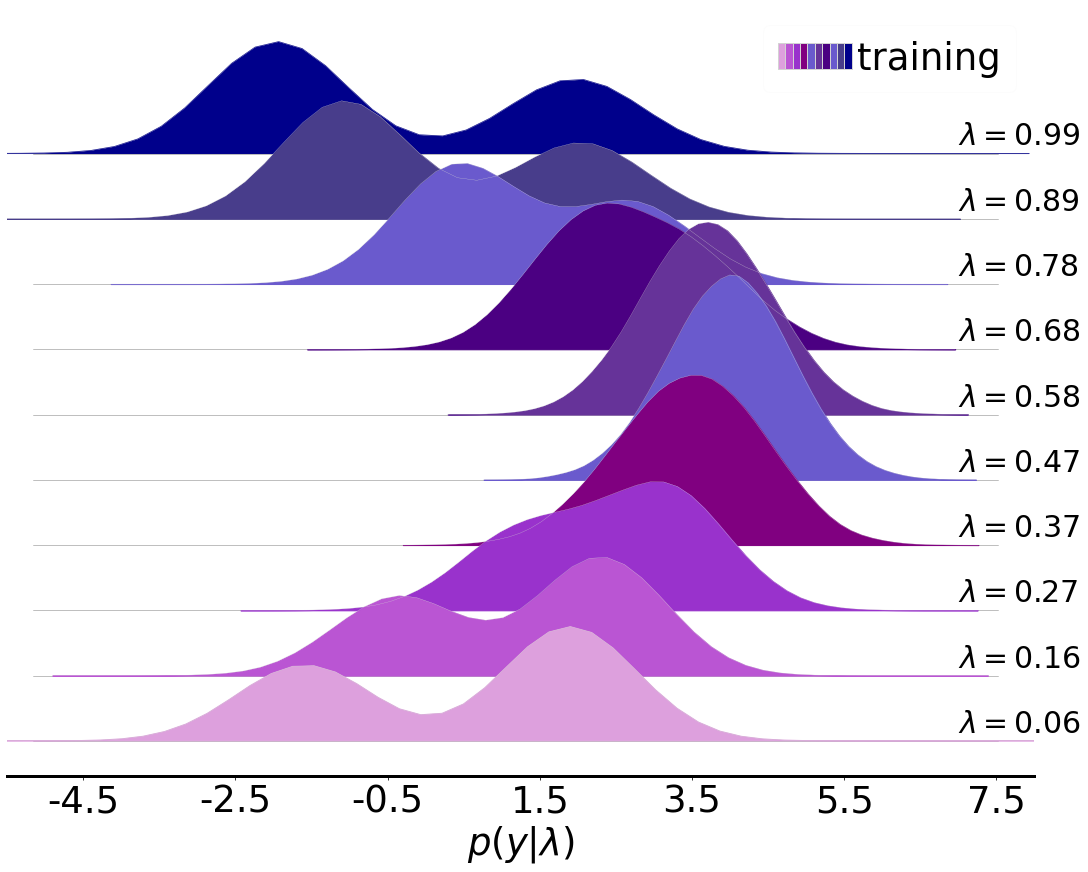}
\caption{Ridge plots demonstrate the true response distribution $p(y|\lambda)$ over a range of $\lambda$-values. The $\lambda$-dependent change in both the shape and location of the response is evident.} 
\label{fig:bimodal} 
\end{figure}

\bt{The intrinsic noise of the previous example is unimodal and only requires first order expansions in the stochastic dimensions to be well represented. We now consider a more complex bimodal example first proposed in~\cite{Zhu:2023} (see Figure~\ref{fig:bimodal}) to demonstrate the versatility of the surrogate construction.}

\bt{Consider a parametrized random variable $y=f(\lambda, \omega)$ with a response distribution in the form of a mixture of two Gaussian PDFs with respective weights $0.4$ and $0.6$,
\begin{align} \label{eq:bimodal}
p(y|\lambda)
	&= 0.5 \phi(1.25 y - c_1(\lambda)) + 0.75 \phi(1.25 y - c_2(\lambda)),
\end{align}
conditioned on parameter $\lambda$. Here $\phi( \cdot)$ denotes the standard normal PDF and the component mean functions are given by $c_1(\lambda) = 5 \sin^2(\pi \lambda) + 5\lambda - 2.5$ and $c_2(\lambda) = 5 \sin^2(\pi \lambda) - 5\lambda + 2.5$ where $\lambda$ is sampled uniformly in $[0,1]$.}

\bt{It was observed by~\cite{Zhu:2023} that the shape and location of the response distribution $p(y|\lambda)$ both depend upon the input parameter $\lambda$. As demonstrated in Figure~\ref{fig:bimodal}, $p(y|\lambda)$ morphs from a bimodal to unimodal distribution as $\lambda$ increases on the subinterval $(0,0.5]$ and then returns to a bimodal distribution over $[0.5, 1)$. Notice that the position of the higher mode switches from the right side of the density at $\lambda=0$ to the left side at $\lambda=1$. At the same time, the mean of $p(y|\lambda)$ increases with $\lambda$ until its maximal value is reached when $\lambda=0.5$.}

\bt{We construct a joint PCE surrogate (see Eq.~\eqref{eq:sequential_pce}) for the response $f(\lambda, \omega)$ that obeys the PDF $p(y|\lambda)$. Representing inverse conditional CDFs of bimodal functions with polynomials is challenging and requires the flexibility given by 15th order terms in the stochastic dimension. Mapping the parametric dimension is comparatively simpler, requiring at most 4th order expansions across all $\bm{z}_{s}(\lambda)$ summarizing quantities. We select the expansion order for each $s$ using model evidence, as before. The surrogate model was constructed using a total of $20$ parametric samples $\lambda$ with $500$ sample replicas per $\lambda$.} 

\begin{figure}[h!]
\centering
\includegraphics[width=0.55\textwidth]{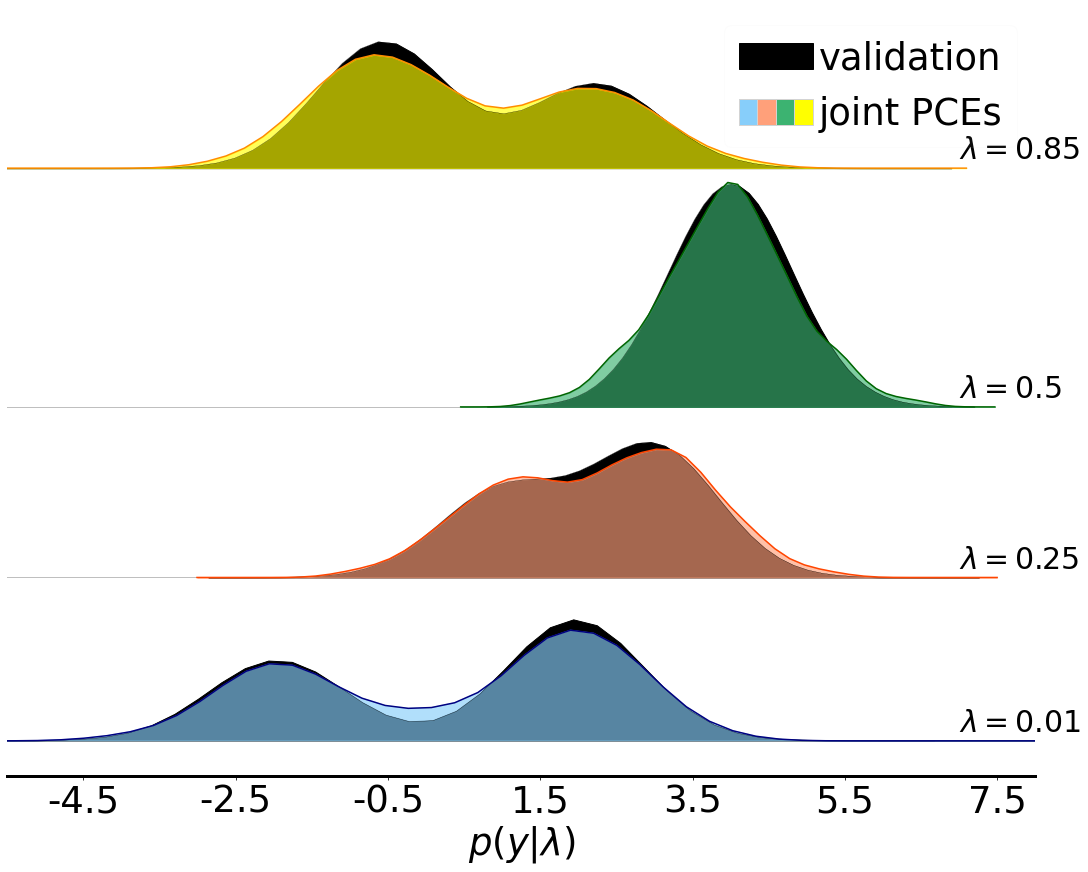}
\caption{Ridge plots demonstrate the joint PCE predicted response distribution $p^{\text{\tiny PCE}}(y|\lambda)$ at parameter locations $\lambda=\{0.01, 0.25, 0.5, 0.85 \}$. } 
\label{fig:bimodal_pce}
\end{figure}

\bt{Distributions predicted by the joint PCE on four validation samples are compared to the true response in Figure~\ref{fig:bimodal_pce}. These samples ($\lambda \in \{0.01, 0.25, 0.5, 0.85\}$) were excluded from the training data and typify the primary shapes the response may take on the domain.
Evinced in Figure~\ref{fig:bimodal_pce}, the joint PCE predictions accurately capture the key $\lambda$-dependent features of the response distribution: change in location, change in distribution shape (bimodal vs unimodal), and change in position of primary mode (from right to left as $\lambda$ increases).}

\section{Conclusion}\label{sec:concl}

We developed a joint PCE model for representing both parametric uncertainty and intrinsic noise of a stochastic model in a compact surrogate form. The construction combines the conventional parametric PCE with a stochastic PCE component constructed with the help of the inverse Rosenblatt transformation. \bt{The latter is commonly known to be challenging for high-dimensional outputs. However, with a limited number of QoIs, we have shown that KDE-based estimation of the associated conditional probabilities produces accurate results. The KDE bandwidth selection, however, remains a challenge and needs to be approached with care.} The joint PCE surrogate is generative as the underlying stochastic germ is represented in a way that can be resampled to generate new realizations of the underlying stochastic model.

 We then extended the joint PCE construction to spatio-temporal random field realizations \bt{employing the KL expansion} and showed that the resulting KLPC surrogate \bt{produces an accurate generative mechanism for the random field. The associated cost is driven by the truncation point of the KL expansion. We found however, that a low number of KL modes are sufficient for acceptable accuracy, and a relatively strong decay of the KL eigenspectrum allows limiting the number of KL terms well below the theoretical limit, which is the actual number of discretized points of the field.} 

The joint PCE \bt{as well as the KLPC} constructions enable analytical extraction of global sensitivity indices via variance decomposition. This allows separating the two sources of uncertainty: intrinsic noise and parametric uncertainty. Decomposing the output variance in this way allows us to assess the impact that intrinsic noise has on the model predictions and can be used to inform the size and quality of the data set used to construct the surrogate. The parametric PCE component of the algorithm, while relatively routine, also includes model selection and basis selection strategies to address the factorial growth of the basis set with the model order and input dimension. 
 
We demonstrated the results on an example heterogeneous catalysis model \bt{and a synthetic Gaussian mixture model with a complex bimodal response distribution}. Improving elements of the surrogate construction workflow, such as removing the parameter independence assumption, investigating alternative bandwidth selection strategies in the Rosenblatt map construction, and implementing a rigorous cross-validation study to complement BCS are subjects of ongoing work. 

\section*{Acknowledgments}
We would like to thank our colleagues Judit Z{\'a}dor and Kyungjoo Kim for their contributions to this work.\\[5pt]
\textbf{Funding}:  This work was done within the Exascale Catalytic Chemistry (ECC) Project, which is supported by the U.S. Department of Energy, Office of Science, Basic Energy Sciences, Chemical Sciences, Geosciences and Biosciences Division, as part of the Computational Chemistry Sciences Program. \\[1pt] This article has been authored by employees of National Technology \& Engineering Solutions of Sandia, LLC under Contract No. DE-NA0003525 with the U.S. Department of Energy (DOE). The employees co-own right, title and interest in and to the article and are responsible for its contents. The United States Government retains and the publisher, by accepting the article for publication, acknowledges that the United States Government retains a non-exclusive, paid-up, irrevocable, world-wide license to publish or reproduce the published form of this article or allow others to do so, for United States Government purposes. The DOE will provide public access to these results of federally sponsored research in accordance with the DOE Public Access Plan https://www.energy.gov/downloads/doe-public-access-plan.

\bibliographystyle{plain}
\bibliography{global}

\end{document}